\PassOptionsToPackage{unicode}{hyperref}
\PassOptionsToPackage{hyphens}{url}
\documentclass[
  american,
]{article}
\usepackage{lmodern}
\usepackage{amssymb,amsmath}
\usepackage{ifxetex,ifluatex}
\ifnum 0\ifxetex 1\fi\ifluatex 1\fi=0 % if pdftex
  \usepackage[T1]{fontenc}
  \usepackage[utf8]{inputenc}
  \usepackage{textcomp} % provide euro and other symbols
\else % if luatex or xetex
  \usepackage{unicode-math}
  \defaultfontfeatures{Scale=MatchLowercase}
  \defaultfontfeatures[\rmfamily]{Ligatures=TeX,Scale=1}
\fi
\IfFileExists{upquote.sty}{\usepackage{upquote}}{}
\IfFileExists{microtype.sty}{% use microtype if available
  \usepackage[]{microtype}
  \UseMicrotypeSet[protrusion]{basicmath} % disable protrusion for tt fonts
}{}
\makeatletter
\@ifundefined{KOMAClassName}{% if non-KOMA class
  \IfFileExists{parskip.sty}{%
    \usepackage{parskip}
  }{% else
    \setlength{\parindent}{0pt}
    \setlength{\parskip}{6pt plus 2pt minus 1pt}}
}{% if KOMA class
  \KOMAoptions{parskip=half}}
\makeatother
\usepackage{xcolor}
\IfFileExists{xurl.sty}{\usepackage{xurl}}{} % add URL line breaks if available
\IfFileExists{bookmark.sty}{\usepackage{bookmark}}{\usepackage{hyperref}}
\hypersetup{
  pdftitle={Approximate leave-future-out cross-validation for Bayesian time series models},
  pdfauthor={Paul-Christian Bürkner \^{}\{1*\}, Jonah Gabry \^{}2, \& Aki Vehtari \^{}1},
  hidelinks,
  pdfcreator={LaTeX via pandoc}}
\usepackage[margin=1in]{geometry}
\usepackage{color}
\usepackage{fancyvrb}

\DefineVerbatimEnvironment{Highlighting}{Verbatim}{commandchars=\\\{\}}
\usepackage{framed}
\definecolor{shadecolor}{RGB}{248,248,248}
\newenvironment{Shaded}{\begin{snugshade}}{\end{snugshade}}

\newcommand{\CommentTok}[1]{\textcolor[rgb]{0.56,0.35,0.01}{\textit{#1}}}

\newcommand{\ControlFlowTok}[1]{\textcolor[rgb]{0.13,0.29,0.53}{\textbf{#1}}}
\newcommand{\DataTypeTok}[1]{\textcolor[rgb]{0.13,0.29,0.53}{#1}}
\newcommand{\DecValTok}[1]{\textcolor[rgb]{0.00,0.00,0.81}{#1}}

\newcommand{\KeywordTok}[1]{\textcolor[rgb]{0.13,0.29,0.53}{\textbf{#1}}}
\newcommand{\NormalTok}[1]{#1}
\newcommand{\OperatorTok}[1]{\textcolor[rgb]{0.81,0.36,0.00}{\textbf{#1}}}

\newcommand{\StringTok}[1]{\textcolor[rgb]{0.31,0.60,0.02}{#1}}

\usepackage{longtable,booktabs}
\usepackage{etoolbox}
\makeatletter
\patchcmd\longtable{\par}{\if@noskipsec\mbox{}\fi\par}{}{}
\makeatother
\IfFileExists{footnotehyper.sty}{\usepackage{footnotehyper}}{\usepackage{footnote}}
\makesavenoteenv{longtable}
\usepackage{graphicx,grffile}
\makeatletter
\def\maxwidth{\ifdim\Gin@nat@width>\linewidth\linewidth\else\Gin@nat@width\fi}
\def\maxheight{\ifdim\Gin@nat@height>\textheight\textheight\else\Gin@nat@height\fi}
\makeatother
\setkeys{Gin}{width=\maxwidth,height=\maxheight,keepaspectratio}
\makeatletter
\def\fps@figure{htbp}
\makeatother
\providecommand{\tightlist}{%
  \setlength{\itemsep}{0pt}\setlength{\parskip}{0pt}}
\usepackage{amsmath}
\usepackage[utf8]{inputenc}
\usepackage[T1]{fontenc}
\usepackage{setspace}
\usepackage{booktabs}
\usepackage{longtable}
\usepackage{array}
\usepackage{multirow}
\usepackage{wrapfig}
\usepackage{float}
\usepackage{colortbl}
\usepackage{pdflscape}
\usepackage{tabu}
\usepackage{threeparttable}
\usepackage{threeparttablex}
\usepackage[normalem]{ulem}
\usepackage{makecell}
\usepackage{xcolor}
\ifxetex
  \usepackage{polyglossia}
  \setmainlanguage[variant=american]{english}
\else
  \usepackage[shorthands=off,main=american]{babel}
\fi
\usepackage[]{natbib}
\title{Approximate leave-future-out cross-validation for Bayesian time series models}
\author{Paul-Christian Bürkner \(^{1*}\), Jonah Gabry \(^2\), \& Aki Vehtari \(^1\)}
\date{\(^1\) Department of Computer Science, Aalto University, Finland\break
\(^2\) Applied Statistics Center and ISERP, Columbia University, USA \break
\(^*\) Corresponding author, Email: \href{mailto:paul.buerkner@gmail.com}{\nolinkurl{paul.buerkner@gmail.com}}}

\begin{document}
\maketitle
\begin{abstract}
One of the common goals of time series analysis is to use the observed series
to inform predictions for future observations. In the absence of any actual
new data to predict, cross-validation can be used to estimate a model's future
predictive accuracy, for instance, for the purpose of model comparison or
selection. Exact cross-validation for Bayesian models is often
computationally expensive, but approximate cross-validation methods have been
developed, most notably methods for leave-one-out cross-validation (LOO-CV).
If the actual prediction task is to predict the future given the past, LOO-CV
provides an overly optimistic estimate because the information from future
observations is available to influence predictions of the past. To properly
account for the time series structure, we can use
leave-future-out cross-validation (LFO-CV). Like exact LOO-CV, exact LFO-CV
requires refitting the model many times to different subsets of the data.
Using Pareto smoothed importance sampling, we propose a method for
approximating exact LFO-CV that drastically reduces the computational costs
while also providing informative diagnostics about the quality of the
approximation.\linebreak\linebreak 
\end{abstract}

\hypertarget{introduction}{%
\section{Introduction}\label{introduction}}

A wide range of statistical models for time series have been developed, finding
applications in industry and nearly all empirical sciences
\citep[e.g., see][]{brockwell2002, hamilton1994}.
One common goal of a time series analysis is to use the observed
series to inform predictions for future time points. In this paper we will
assume a Bayesian approach to time series modeling, in which case if it is
possible to sample from the posterior \emph{predictive} distribution implied by a
given time series model, then it is straightforward to generate predictions as
far into the future as we want. When working in discrete time we will refer to
the task of predicting a sequence of \(M\) future observations as \(M\)-step-ahead
prediction (\(M\)-SAP).

It is easy to evaluate the \(M\)-SAP performance of a time series
model by comparing the predictions to the observed sequence of \(M\) future data
points once they become available. However, we would often like to estimate
the future predictive performance of a model \emph{before} we are able to collect
additional observations. If there are many competing models we may also need to
first decide which model (or which combination of the models) to
rely on for prediction \citep{geisser1979, hoeting1999, vehtari2002, ando2010, vehtari2012}.

In the absence of new data, one general approach for evaluating a model's
predictive accuracy is cross-validation. The data is first split into two
subsets, then we fit the statistical model to the first subset and evaluate
predictive performance with the second subset. We may do this once or many
times, each time leaving out a different subset.

If the data points are not ordered in time, or if the goal is to assess the
non-time-dependent part of the model, then we can use leave-one-out
cross-validation (LOO-CV). For a data set with \(N\) observations, we refit the
model \(N\) times, each time leaving out one of the \(N\) observations and assessing
how well the model predicts the left-out observation. Due to the number of
required refits, exact LOO-CV is computationally expensive, in particular when
performing full Bayesian inference and refitting the model means estimating a
new posterior distribution rather than a point estimate. But it is possible to
approximate exact LOO-CV using Pareto smoothed importance sampling
\citep[PSIS;][]{vehtari2017loo, vehtari2019psis}.
PSIS-LOO-CV only requires a single fit of the full model and has sensitive
diagnostics for assessing the validity of the approximation.

However, using LOO-CV with times series models is problematic if the goal is to
estimate the predictive performance for future time points. Leaving out only one
observation at a time will allow information from the future to influence
predictions of the past (i.e., data from times \(t+1, t+2, \ldots,\) would inform
predictions for time \(t\)). Instead, to apply the idea of cross-validation to the
\(M\)-SAP case we can use what we will refer to as leave-\emph{future}-out
cross-validation (LFO-CV). LFO-CV does not refer to one particular prediction
task but rather to various possible cross-validation approaches that all involve
some form of prediction of future time points.

Like exact LOO-CV, exact LFO-CV requires refitting the model many times to
different subsets of the data, which is computationally expensive, in particular
for full Bayesian inference. In this paper, we extend the ideas from PSIS-LOO-CV
and present PSIS-LFO-CV, an algorithm that typically only requires refitting a
time series model a small number times. This will make LFO-CV tractable for many
more realistic applications than previously possible, including time series
model averaging using stacking of predictive distributions \citep{yao2018}.

The efficiency of PSIS-LFO-CV compared to exact LFO-CV relies on the
ability to compute samples from the posterior predictive distribution (required
for the importance sampling) in much less time than it takes to fully refit the
model. This assumption is most likely justified when estimating a model using
full Bayesian inference via MCMC, variational inference, or related methods as
they are very computationally intensive. We do not make any
assumptions about \emph{how} samples from the posterior or the posterior predictive
density at a given point in time have been obtained.

Our proposed algorithm was motivated by the practical need for efficient
cross-validation tools for Bayesian time series models fit using generic
probabilistic programming frameworks, such as Stan \citep{carpenter2017}, JAGS
\citep{jags}, PyMC3 \citep{pymc3} and Pyro \citep{pyro}. These frameworks have become very
popular in recent years also for analysis of time series models. For many
models, inference could also be performed using sequential Monte Carlo (SMC)
\citep[e.g.,][]{doucet2000, andrieu2010} using, for example, the SMC-specific framework
Birch \citep{birch}. The implementation details of LFO-CV for SMC algorithms are
beyond the scope of this paper.\footnote{Most SMC algorithms use importance
sampling and LFO-CV could be obtained as a by-product, with computation
resembling the approach we present here. The proposal distribution at each step
and the applied "refit" approach (when the importance sampling weights become
degenerate) are specific to each SMC algorithm.}

The structure of the paper is as follows. In Section \ref{m-sap}, we introduce
the idea and various forms of \(M\)-step-ahead prediction and how to approximate
them using PSIS. In Section \ref{simulations}, we evaluate the accuracy of the
approximation using extensive simulations. Then, in Section \ref{case-studies},
we provide two case studies demonstrating the application of PSIS-LFO-CV methods
to real data sets. In the first we model changes in the water level of Lake
Huron and in the second the date of the yearly cherry blossom in Kyoto. We end
in Section \ref{discussion} with a discussion of the usefulness and limitations
of our approach.

\hypertarget{m-sap}{%
\section{\texorpdfstring{\(M\)-step-ahead predictions}{M-step-ahead predictions}}\label{m-sap}}

Assume we have a time series of observations \(y = (y_1, y_2, \ldots, y_N)\)
and let \(L\) be the \emph{minimum} number of observations from the series that
we will require before making predictions for future data. Depending on the
application and how informative the data are, it may not be possible to make
reasonable predictions for \(y_{i+1}\) based on \((y_1, \dots, y_{i})\) until \(i\) is
large enough so that we can learn enough about the time series to predict future
observations. Setting \(L=10\), for example, means that we will only assess
predictive performance starting with observation \(y_{11}\), so that we
always have at least 10 previous observations to condition on.

In order to assess \(M\)-SAP performance we would like to compute the
predictive densities

\begin{equation}
p(y_{i+1:M} \,|\, y_{1:i}) = 
  p(y_{i+1}, \ldots, y_{i + M} \,|\, y_{1},...,y_{i}) 
\end{equation}

for each \(i \in \{L, \ldots, N - M\}\), where we use \(y_{1:i} = (y_{1}, \ldots, y_{i})\)
and \(y_{i+1:M} = y_{(i+1):(i+M)} = (y_{i+1}, \ldots, y_{i + M})\)
to shorten the notation\footnote{Note that the here-used ``\(:\)'' operator has precedence over the
  ``\(+\)'' operator following the R programming language definition.}. As a global measure of predictive accuracy, we
can use the expected log predictive density \citep[ELPD;][]{vehtari2017loo}, which,
for M-SAP, can be defined as

\begin{equation}
\label{ELPD}
{\rm ELPD} = \sum_{i=L}^{N - M} 
  \int p_t(\tilde{y}_{i+1:M}) \log p(\tilde{y}_{i+1:M} \,|\, y_{1:i})
  \, {\rm d} \, \tilde{y}_{i+1:M}.
\end{equation}

The distribution \(p_t(\tilde{y}_{i+1:M})\) describes the true data generating
process for new data \(\tilde{y}_{i+1:M}\). As these true data generating
processes are unknown, we approximate the ELPD using LFO-CV of the observed
responses \(y_{i+1:M}\), which constitute a particular realization of
\(\tilde{y}_{i+1:M}\). This approach of approximationg the true data generating
process with observed data is fundamental to all cross-validation procedures.
As we have no direct access to the underlying truth, the error implied by
this approximation is hard to quantify but also unavoidable \citep[c.f.,][]{bernardo1994}.

Plugging in the realization \(y_{i+1:M}\) for \(\tilde{y}_{i+1:M}\) leads
to \citep[c.f.,][]{bernardo1994, vehtari2012}:

\begin{equation}
{\rm ELPD}_{\rm LFO} = \sum_{i=L}^{N - M} \log p(y_{i+1:M} \,|\, y_{1:i}).
\end{equation}

The quantities \(p(y_{i+1:M} \,|\, y_{1:i})\) can be computed with the help of the
posterior distribution \(p(\theta \,|\, y_{1:i})\) of the parameters \(\theta\)
conditional on only the first \(i\) observations of the time series:

\begin{equation}
\label{Lpred}
p(y_{i+1:M} \,| \, y_{1:i}) = 
  \int p(y_{i+1:M} \,| \, y_{1:i}, \theta) \, 
    p(\theta\,|\,y_{1:i}) \, {\rm d} \theta. 
\end{equation}

In order to evaluate predictive performance of future data, it is important to
predict \(y_{i+1:M}\) only conditioning on the past data \(y_{1:i}\) and not on the
present data \(y_{i+1:M}\). Including the present data in the posterior
estimation, that is, using the posterior \(p(\theta\,|\,y_{1:(i+M)})\) in
\eqref{Lpred}, would result in evaluating in-sample fit. This corresponds to
what \citet{vehtari2017loo} call \emph{log-predictive density} (LPD), which overestimates
predictive performance for future data \citep{vehtari2017loo}.

Most time series models do not have conditionally independent observations, that
is, \(y_{i+1:M}\) depend on \(y_{1:i}\) even after conditioning on \(\theta\). As
such, we cannot simplify the integrand in (\ref{Lpred}) and always need to take
\(y_{1:i}\) into account when computing the predictive density of \(y_{i+1:M}\). The
concept of conditional independence of observations is closely related to the
concept of factorizability of likelihoods. For the purpose of LFO-CV, we can
safely use the time-ordering naturally present in time-series data to obtain a
factorized likelihood even if observations are not conditionally independent.
See \citet{buerkner:non-factorizable} for discussion on computing predictive densities
of non-factorized models and factorizability in general.

In practice, we will not be able to directly solve the integral in
(\ref{Lpred}), but instead have to use Monte-Carlo methods to approximate it.
Having obtained \(S\) random draws \((\theta_{1:i}^{(1)}, \ldots, \theta_{1:i}^{(S)})\)
from the posterior distribution \(p(\theta\,|\,y_{1:i})\), we can estimate
\(p(y_{i+1:M} | y_{1:i})\) as

\begin{equation}
p(y_{i+1:M} \,|\, y_{1:i}) \approx \frac{1}{S}
\sum_{s=1}^S p(y_{i+1:M} \,|\, y_{1:i}, \theta_{1:i}^{(s)}).
\end{equation}

In this paper we use ELPD as a measure of predictive accuracy, but \(M\)-SAP (and
the approximations we introduce below) may also be based on other global
measures of accuracy such as the root mean squared error (RMSE) or the median
absolute deviation (MAD). The reason for our focus on ELPD is that it evaluates
a distribution rather than a point estimate (like the mean or median) to provide
a measure of out-of-sample predictive performance, which we see as favorable
from a Bayesian perspective \citep{vehtari2012}. The code we provide on GitHub
(\url{https://github.com/paul-buerkner/LFO-CV-paper}) is modularized to support
arbitrary measures of accuracy as long as they can be represented in a pointwise
manner, that is, as increments per observation. In Appendix C we also provide
additional simulation results using RMSE instead of ELPD.

\hypertarget{approximate-MSAP}{%
\subsection{\texorpdfstring{Approximate \(M\)-step-ahead predictions}{Approximate M-step-ahead predictions}}\label{approximate-MSAP}}

The equations above make use of posterior distributions from many
different fits of the model to different subsets of the data. To obtain
the predictive density \(p(y_{i+1:M} \,|\, y_{1:i})\), a model is fit to
only the first \(i\) data points, and we need to do this for every value of
\(i\) under consideration (all \(i \in \{L, \ldots, N - M\}\)).
Below, we present a new algorithm to reduce the number of models that need
to be fit for the purpose of obtaining each of the densities
\(p(y_{i+1:M} \,|\, y_{1:i})\). This algorithm relies in a central manner on
Pareto smoothed importance sampling \citep{vehtari2017loo, vehtari2019psis}, which
we will briefly review next.

\hypertarget{psis}{%
\subsubsection{Pareto smoothed importance sampling}\label{psis}}

Importance sampling is a technique for computing expectations with
respect to some target distribution using an approximating proposal distribution
that is easier to draw samples from than the actual target. If \(f(\theta)\) is
the target and \(g(\theta)\) is the proposal distribution, we can write any
expectation of some function \(h(\theta)\) with respect to \(f\) as

\begin{equation}
\mathbb{E}_f[h(\theta)] = \int h(\theta) f(\theta) \,d\, \theta 
 = \frac{\int [h(\theta) f(\theta) / g(\theta)] g(\theta) \,d\, \theta}
    {\int [f(\theta) / g(\theta)] g(\theta) \,d\, \theta} 
 = \frac{\int h(\theta) r(\theta) g(\theta) \,d\, \theta}
    {\int r(\theta) g(\theta) \,d\, \theta}
\end{equation}

with importance ratios
\begin{equation}
r(\theta) = \frac{f(\theta)}{g(\theta)}.
\end{equation}

Accordingly, if \(\theta^{(s)}\) are \(S\) random draws from \(g(\theta)\), we can
approximate

\begin{equation}
\mathbb{E}_f[h(\theta)] \approx 
\frac{\sum_{s=1}^S h(\theta^{(s)}) r(\theta^{(s)})}{\sum_{s=1}^S r(\theta^{(s)})},
\end{equation}

provided that we can compute the raw importance ratios \(r(\theta^{(s)})\) up to
some multiplicative constant. The raw importance ratios serve as weights on the
corresponding random draws in the approximation of the quantity of interest.

The main problem with this approach is that the raw importance ratios tend to
have large or infinite variance and results can be highly unstable. In order to
stabilize the computations, we can perform the additional step of regularizing
the largest raw importance ratios using the corresponding quantiles of the
generalized Pareto distribution fitted to the largest raw importance ratios.
This procedure is called Pareto smoothed importance sampling
\citep[PSIS;][]{vehtari2017loo, vehtari2019psis, loo2019} and has been demonstrated to
have a lower error and faster convergence rate than other commonly used
regularization techniques \citep{vehtari2019psis}.

In addition, PSIS comes with a useful diagnostic to evaluate
the quality of the importance sampling approximation. The shape parameter \(k\)
of the generalized Pareto distribution fit to the largest importance ratios provides
information about the number of existing moments of the distribution of the
weights (smoothed ratios) and the actual importance sampling estimate.
When \(k<0.5\), the weight distribution has finite variance and the central limit
theorem ensures fast convergence of the importance sampling estimate with
increasing number of draws. This implies that approximate LOO-CV via PSIS is
highly accurate for \(k<0.5\) \citep{vehtari2019psis}. For \(0.5 \leq k < 1\), a
generalized central limit theorem holds, but the convergence rate drops quickly
as \(k\) increases \citep{vehtari2019psis}. Using both mathematical analysis
and numerical experiments, PSIS has been shown to be
relatively robust for \(k < 0.7\) \citep{vehtari2017loo, vehtari2019psis}. As such,
the default threshold is set to \(0.7\) when performing
PSIS LOO-CV \citep{vehtari2017loo, loo2019}.

\hypertarget{psis-MSAP}{%
\subsubsection{\texorpdfstring{PSIS applied to \(M\)-step-ahead predictions}{PSIS applied to M-step-ahead predictions}}\label{psis-MSAP}}

We now turn back to the task of estimating \(M\)-step-ahead performance for
time series models. First, we refit the model using the first \(L\) observations
of the time series and then perform a single exact \(M\)-step-ahead prediction
step for \(p(y_{L+1:M} \,|\, y_{1:L})\) as per \eqref{Lpred}.
Recall that \(L\) is the minimum number of observations we have deemed
acceptable for making predictions (setting \(L=0\) means the first data point will
be predicted only based on the prior). We define \(i^\star = L\) as the current
point of refit. Next, starting with \(i = i^\star + 1\), we
approximate each \(p(y_{i+1:M} \,|\, y_{1:i})\) via

\begin{equation}
 p(y_{i+1:M} \,|\, y_{1:i}) \approx
   \frac{ \sum_{s=1}^S w_i^{(s)}\, p(y_{i+1:M} \,|\, y_{1:i}, \theta^{(s)})}
        { \sum_{s=1}^S w_i^{(s)}},
\end{equation}

where \(\theta^{(s)} = \theta^{(s)}_{1:i^\star}\) are draws from the
posterior distribution based on the first \(i^\star\) observations
and \(w_i^{(s)}\) are the PSIS weights obtained in two steps.
First, we compute the raw importance ratios

\begin{equation}
r_i^{(s)} = r_i(\theta^{(s)}) = 
\frac{f_{1:i}(\theta^{(s)})}{f_{1:i^\star}(\theta^{(s)})} 
\propto \frac{
p(\theta^{(s)}) \prod_{j \in 1:i} p(y_j \,|\, y_{1:(j-1)}, \theta^{(s)}) 
}{
p(\theta^{(s)}) \prod_{j \in 1:i^\star} p(y_j \,|\, y_{1:(j-1)}, \theta^{(s)}) 
}
= \prod_{j \in (i^\star + 1):i} p(y_j \,|\, y_{1:(j-1)}, \theta^{(s)}),
\end{equation}

and then stabilize them using PSIS as described in Section \ref{psis}. The
function \(f_{1:i}\) denotes the posterior distribution based on the first \(i\)
observations, that is, \(f_{1:i} = p(\theta \,|\, y_{1:i})\), with \(f_{1:i^\star}\)
defined analogously. The index set \((i^\star + 1):i\) indicates all observations
which are part of the data for the model \(f_{1:i}\) whose predictive performance
we are trying to approximate but not for the actually fitted model
\(f_{1:i^\star}\). The proportional statement arises from the fact that
we ignore the normalizing constants \(p(y_{1:i})\) and \(p(y_{1:i^\star})\)
of the compared posteriors, which leads to a self-normalized variant of
PSIS \citep[c.f.][]{vehtari2017loo}.

\begin{figure}
\centering
\includegraphics{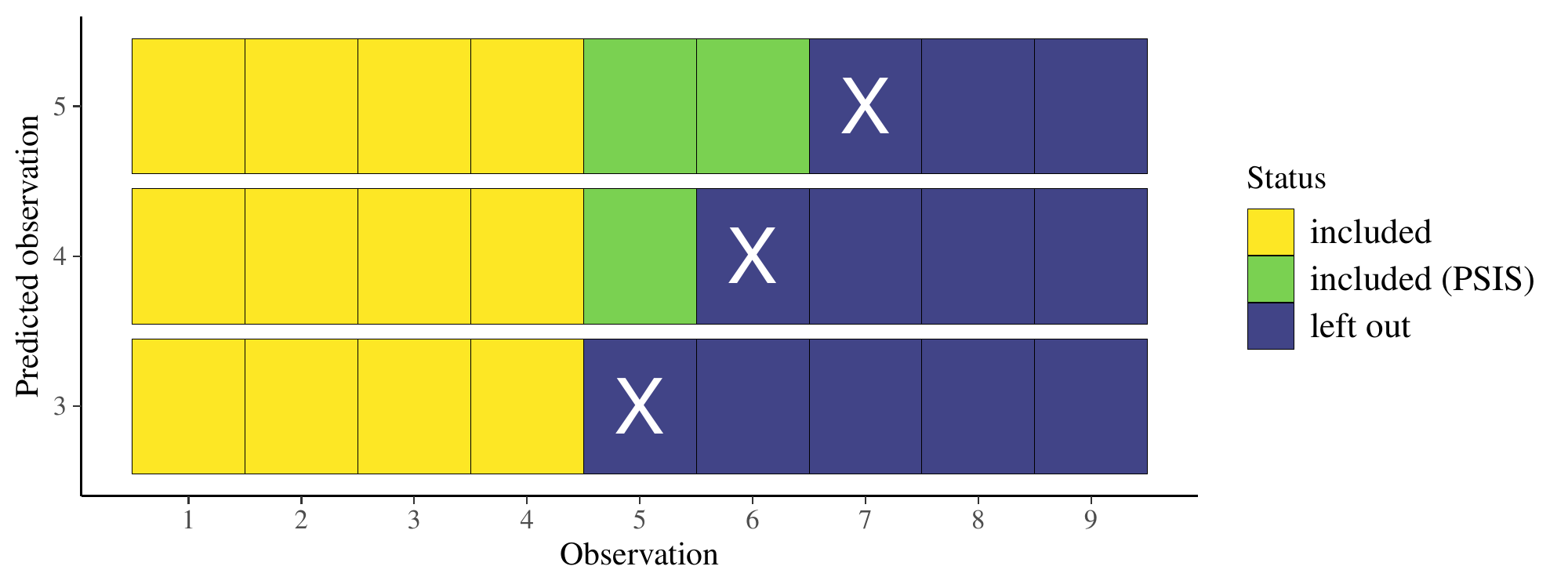}
\caption{\label{fig:vis-msap}Visualisation of PSIS approximated one-step-ahead predictions. Predicted observations are indicated by \textbf{X}. In the shown example, the model was last refit at the \(i^\star = 4\)th observation.}
\end{figure}

Continuing with the next observation, we gradually increase \(i\) by \(1\) (we move
forward in time) and repeat the process. At some observation \(i\), the
variability of the importance ratios \(r_i^{(s)}\) will become too large and
importance sampling will fail. We will refer to this particular value of \(i\) as
\(i^\star_1\). To identify the value of \(i^\star_1\), we check for which value of
\(i\) does the estimated shape parameter \(k\) of the generalized Pareto
distribution first cross a certain threshold \(\tau\) \citep{vehtari2019psis}. Only
then do we refit the model using the observations up to \(i^\star_1\) and restart
the process from there by setting \(\theta^{(s)} = \theta^{(s)}_{1:i^\star_1}\)
and \(i^\star = i^\star_1\) until the next refit.
An illustration of this procedure is shown in Figure
\ref{fig:vis-msap}.

This bears some resemblance to Sequential Monte Carlo, also known as
particle or Monte Carlo filtering \citep[e.g.,][]{gordon1993, kitagawa1996, doucet2000, andrieu2010}, in that applying SMC to state space models also
entails moving forward in time and using importance sampling to approximate the
next state from the information in the previous states \citep{kitagawa1996, andrieu2010}. However, in our case we are assuming we can sample from the
posterior distribution and are instead concerned with estimating the model's
predictive performance. Unlike SMC, PSIS-LFO-CV also entails a full recomputation
of the model via Markov chain Monte Carlo (MCMC) once importance sampling fails.

In some cases we may only need to refit once and in other cases we will find a
value \(i^\star_2\) that requires a second refitting, maybe an \(i^\star_3\) that
requires a third refitting, and so on. We refit as many times as is required
(only when \(k > \tau\)) until we arrive at observation \(i = N - M\). A detailed
description of the algorithm in the form of pseudo code is provided in Appendix
A. If the data are comprised of multiple \emph{independent} time series, the algorithm can
be applied to each of the time series separately and the resulting ELPD values
can be summed up afterwards. If the data are comprised of multiple \emph{dependent}
time series, the algorithm should be applied to the joint likelihood across all
time-series for each observation \(i\) in order to take the dependency into
account.

Instead of moving forward in time, it is also possible to do
PSIS-LFO-CV moving backwards in time. However, in that case the target posterior
is approximated by a distribution based on more observations and, as such,
the proposal distribution is narrower than the target. This can result in
highly influential importance weights more often than when
the proposal is wider than the target, which is the case
for the forward PSIS-LFO-CV described above. In Appendix B, we show
that moving backwards indeed requires more refits than moving forward, and
without any increase in accuracy. When we refer to the PSIS-LFO-CV algorithm
in the main text we are referring to the forward version.

The threshold \(\tau\) is crucial to the accuracy and speed of the PSIS-LFO-CV
algorithm. If \(\tau\) is too large then we need fewer refits but accuracy will
suffer. If \(\tau\) is too small then accuracy will be higher but many refits will
be required and the computation time may be impractical. Limiting the number of
refits without sacrificing too much accuracy is essential since almost all of
the computation time for exact cross-validation of Bayesian models is spent
fitting the models (not calculating the predictions). Therefore, any reduction
we can achieve in the number of refits essentially implies a proportional
reduction in the overall time required for cross-validation of Bayesian models.
We will come back to the issue of setting appropriate thresholds in Section
\ref{simulations}.

\hypertarget{simulations}{%
\section{Simulations}\label{simulations}}

\begin{figure}
\centering
\includegraphics{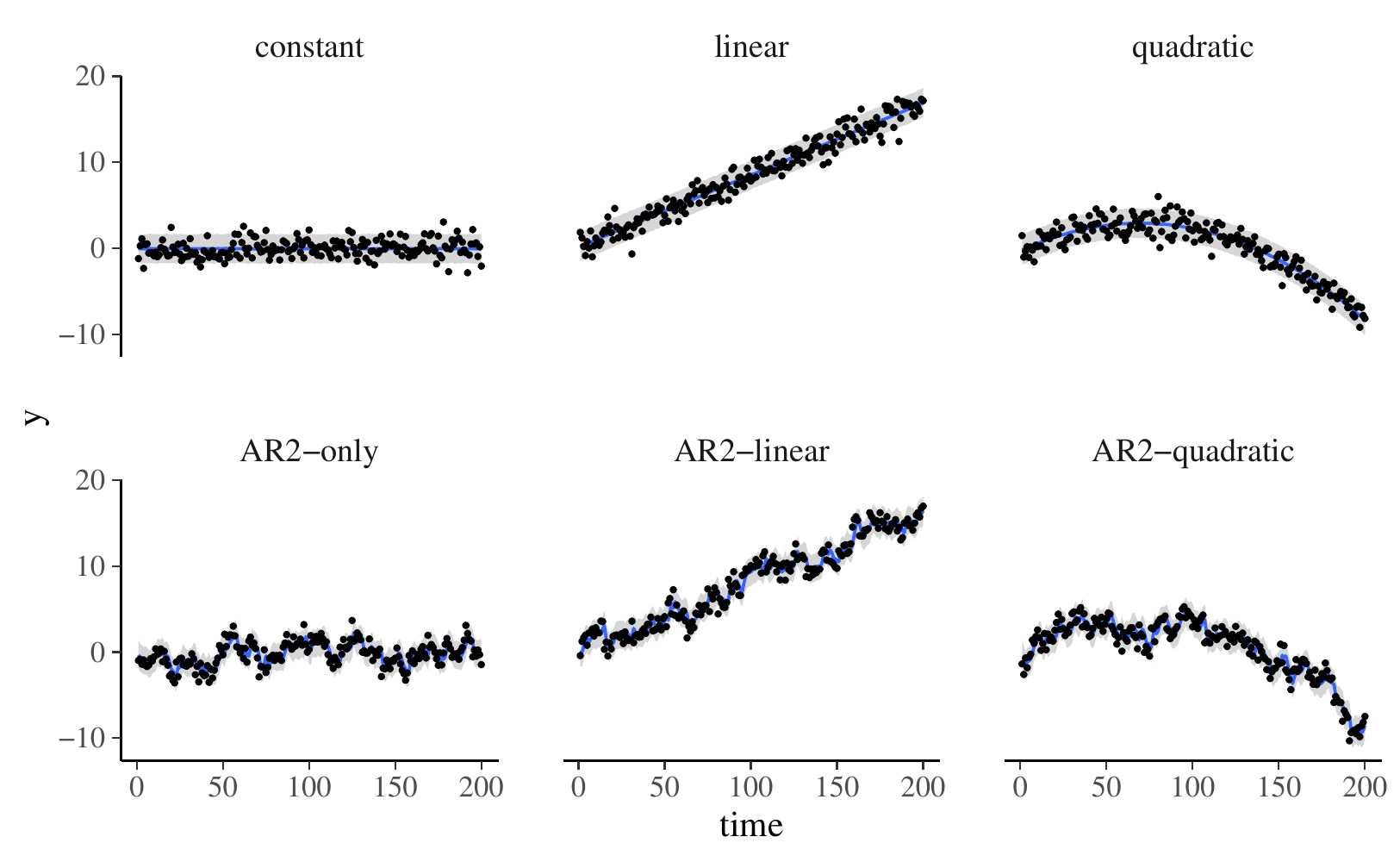}
\caption{\label{fig:simmodels}Illustration of the models used in the simulations. Black points are observed data. The blue line represents posterior predictions of the model resembling the true data-generating process with 90\% prediction intervals shown in gray. More details are provided in the text.}
\end{figure}

To evaluate the quality of the PSIS-LFO-CV approximation, we performed a
simulation study in which the following conditions were systematically varied:

\begin{itemize}
\tightlist
\item
  The number \(M\) of future observations to be predicted took on values of \(M = 1\) and \(M = 4\).
\item
  The threshold \(\tau\) of the Pareto \(k\) estimates was varied between \(k = 0.5\)
  to \(k = 0.7\) in steps of \(0.1\).
\item
  Six different data generating models were evaluated, with linear and/or
  quadratic terms and/or autoregressive terms of order 2 (see Figure
  \ref{fig:simmodels} for an illustration).
\end{itemize}

In all cases the time series consisted of \(N = 200\) observations and the minimal
number of observations required before make predictions was set to \(L = 25\).
We ran \(100\) simulation trials per condition.

Autoregressive (AR) models are some of the most commonly used time series models.
An AR(p) model -- an autoregressive model of order \(p\) -- can be defined as

\begin{equation}
y_i = \eta_i + \varepsilon_i \quad \text{with} \quad
\varepsilon_i = \sum_{k = 1}^p \varphi_k \varepsilon_{i - k} + e_i,
\end{equation}
where \(\eta_i\) is the linear predictor for the \(i\)th observation, \(\varphi_k\)
are the autoregressive parameters on the residuals \(\varepsilon_i\), and \(e_i\) are
pairwise independent errors, which are usually assumed to be normally
distributed with equal variance \(\sigma^2\). The model implies a recursive
formula that allows for computing the right-hand side of the equation for
observation \(i\) based on the values of the equations computed for previous
observations. Observations from an AR process are therefore not conditionally
independent by definition, but the likelihood still factorizes because we
can write down a separate contribution for each observation \citep[see][ for more discussion on factorizability of statistical
models]{buerkner:non-factorizable}.

In the quadratic model with AR(2) effects (the most complex model in our
simulations), the true data generating process was defined as
\begin{equation}
y_i = b_0 + b_1 t + b_2 t^2 + \varepsilon_i \quad \text{with} \quad
\varepsilon_i = \varphi_1 \varepsilon_{i - 1} + \varphi_2 \varepsilon_{i - 2} + e_i,
\end{equation}
where \(t\) is the time variable scaled to the unit interval, that is, \(t = 0\) for
the smallest time point (\(1\) in our simulations) and \(t = 1\) for the largest
time point (\(200\) in our simulations). Specifically, we set the true regression
coefficients to the values of \(b_0 = 0\), \(b_1 = 17\), \(b_2 = 25\), and the true
autocorrelation parameters to \(\varphi_1 = 0.5\), and \(\varphi_2 = 0.3\) (see
Figure \ref{fig:simmodels} for an illustration). The choices of the regression
coefficients were done so that neither the linear nor quadratic term dominates
the other within the specified time frame. The values of the autocorrelation
parameters were set to represent typical positively autocorrelated data.
In the simulation conditions without linear and/or quadratic and/or AR(2) terms,
the corresponding true parameters were simply fixed to zero. We always fit the
true data generating model to the data. This is neither required for the
validity of LFO-CV in general nor for the validity of the comparison between
exact and approximate versions but simply a choice of convenience. For example,
a linear model without autocorrelation is used when all but \(b_0\) and \(b_1\) were
set to zero in the simulations.

In addition to exact and approximate LFO-CV, we also computed approximate LOO-CV
for comparison. This is not because we think LOO-CV is a generally appropriate
approach for time series models, but because, in the absence of any approximate
LFO-CV method, researchers may have used approximate LOO-CV for time series
models in the past simply because it was the only available option.
Demonstrating that LOO-CV is a biased estimate of LFO-CV underscores
the importance of developing methods better suited for the task.

All simulations were done in R \citep{R2018} using the brms package \citep{brms1, brms2} together with the probabilistic programming language Stan
\citep{carpenter2017, rstan2019} for model fitting, the loo R package
\citep{loo2019} for the PSIS computations, and several tidyverse R packages
\citep{tidyverse} for data processing. The full code and all results are available on
Github (\url{https://github.com/paul-buerkner/LFO-CV-paper}).

\hypertarget{sim_results}{%
\subsection{Results}\label{sim_results}}

In this section we focus on the ELPD as a measure out-of-sample predictive
performance for reasons outlined in Section \ref{m-sap}. In Appendix C,
we provide additional simulation results for the RMSE.

\begin{figure}
\centering
\includegraphics{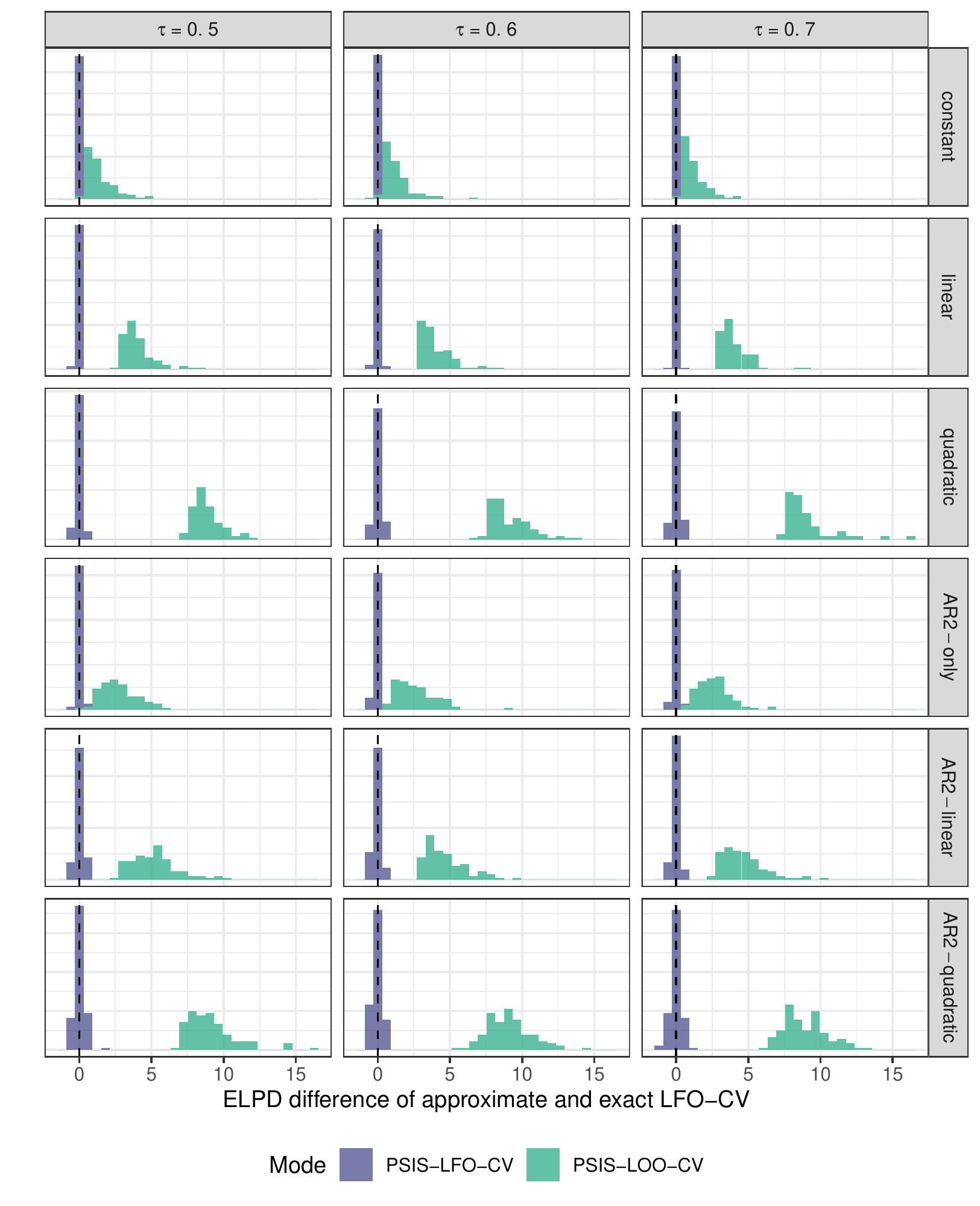}
\caption{\label{fig:1sap}Simulation results of 1-step-ahead predictions. Histograms are based on 100 simulation trials of time series with \(N = 200\) observations requiring at least \(L = 25\) observations to make predictions. The black dashed lines indicates the exact LFO-CV result.}
\end{figure}

Results of the 1-SAP simulations are visualized in Figure \ref{fig:1sap}.
Comparing the columns of Figure \ref{fig:1sap}, it is clearly visible that the
the accuracy of the PSIS approximation is independent of the threshold \(\tau\)
when \(\tau\) is within the interval \([0.5,0.7]\) motivated in \ref{psis} \citep[this
would not be the case if \(\tau\) was allowed to be larger;][]{vehtari2019psis}. For
all conditions, the PSIS-LFO-CV approximation is highly accurate, that is, both
unbiased and low in variance around the corresponding exact LFO-CV value
(represented by the dashed line in Figure \ref{fig:1sap}). The proportion of
observations at which refitting the model was required did not exceed \(3\%\)
under any of the conditions and only increased minimally when decreasing \(\tau\)
(see Table \ref{tab:refits}). At least for the models investigated in our
simulations, using \(\tau = 0.7\) seems to be sufficient for achieving high
accuracy and as such there is no need to lower the threshold below that value.
As expected, LOO-CV (the lighter histograms in Figure \ref{fig:1sap}) is a
biased estimate of the 1-SAP performance for all non-constant models, in
particular for models with a trend in the time series. More precisely, LOO-CV is
positively biased, which implies that it systematically overestimates \(M\)-SAP
performance of time series models.

\begin{table}

\caption{\label{tab:refits}Mean proportions of required refits for PSIS-LFO-CV.}
\centering
\begin{threeparttable}
\begin{tabular}[t]{lrrrrrrr}
\toprule
M & $\tau$ & constant & linear & quadratic & AR2-only & AR2-linear & AR2-quadratic\\
\midrule
1 & 0.5 & 0.01 & 0.01 & 0.02 & 0.01 & 0.02 & 0.03\\
 & 0.6 & 0.01 & 0.01 & 0.02 & 0.01 & 0.02 & 0.02\\
 & 0.7 & 0.01 & 0.01 & 0.02 & 0.01 & 0.01 & 0.02\\
4 & 0.5 & 0.01 & 0.01 & 0.02 & 0.01 & 0.02 & 0.03\\
 & 0.6 & 0.01 & 0.01 & 0.02 & 0.01 & 0.02 & 0.02\\
\addlinespace
 & 0.7 & 0.01 & 0.01 & 0.02 & 0.01 & 0.01 & 0.02\\
\bottomrule
\end{tabular}
\begin{tablenotes}
\item Note: Results are based on 100 simulation trials of time series with $N = 200$ observations requiring at least $L = 25$ observations to make predictions. Abbreviations: $\tau$ = threshold of the Pareto $k$ estimates; $M$ = number of predicted future observations.
\end{tablenotes}
\end{threeparttable}
\end{table}

\begin{figure}
\centering
\includegraphics{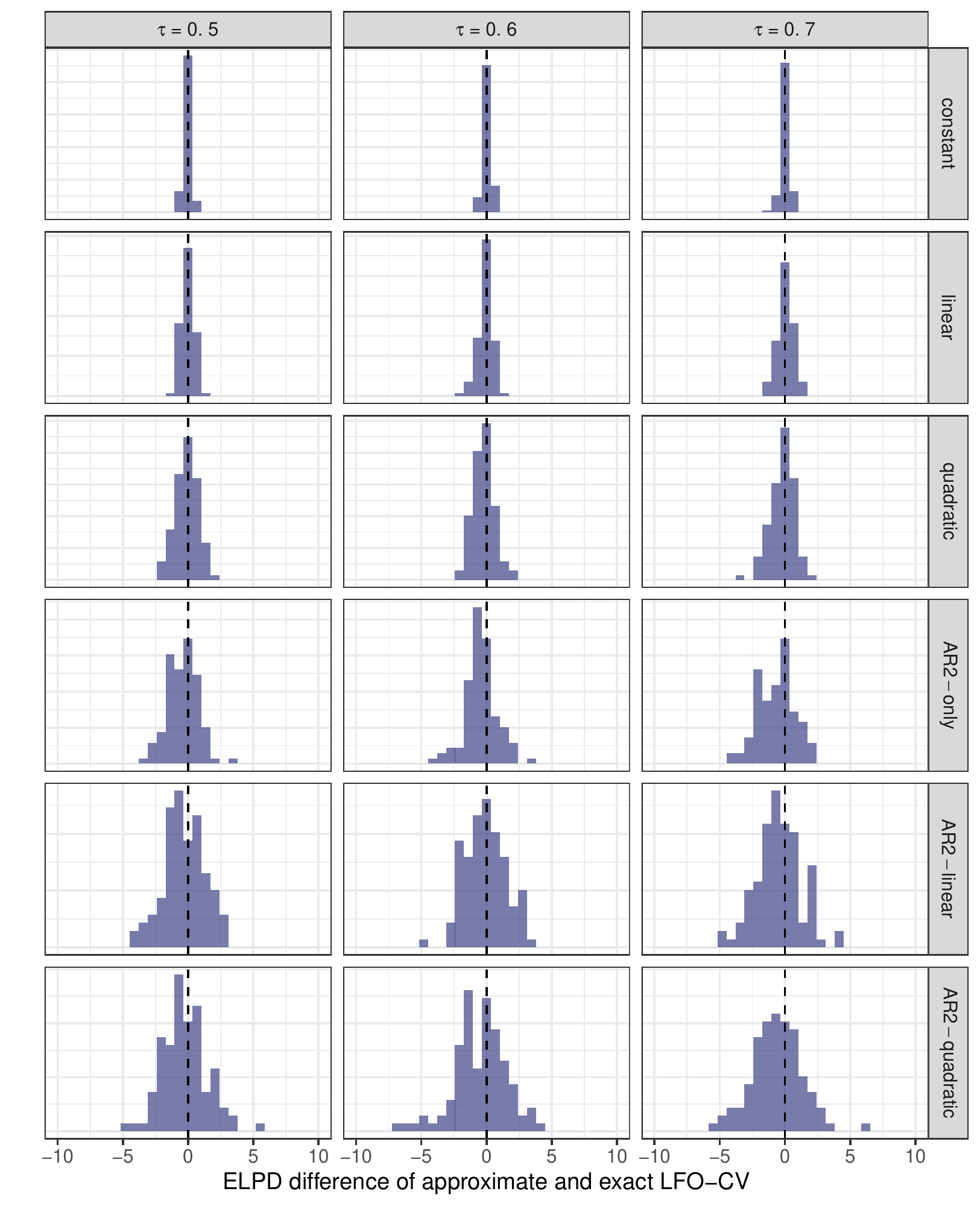}
\caption{\label{fig:4sap}Simulation results of 4-step-ahead predictions. Histograms are based on 100 simulation trials of time series with \(N = 200\) observations requiring at least \(L = 25\) observations to make predictions. The black dashed lines indicates the exact LFO-CV result.}
\end{figure}

Results of the 4-SAP simulations are visualized in Figure \ref{fig:4sap}.
Comparing the columns of Figure \ref{fig:4sap}, it is again clearly visible
that the accuracy of the PSIS approximation is independent of the threshold
\(\tau\). The proportion of observations at which refitting the model was required
did not exceed \(3\%\) under any condition and only increased minimally when
decreasing \(\tau\) (see Table \ref{tab:refits}). In light of the corresponding
1-SAP results presented above, this is not surprising because the procedure for
determining the necessity of a refit is independent of \(M\) (see Section
\ref{approximate-MSAP}). PSIS-LOO-CV is not displayed in Figure \ref{fig:4sap} as
the number of observations predicted at each step (4 vs.~1) makes 4-SAP LFO-CV
and LOO-CV incomparable.

\hypertarget{case-studies}{%
\section{Case Studies}\label{case-studies}}

\hypertarget{case-LH}{%
\subsection{Annual measurements of the level of Lake Huron}\label{case-LH}}

\begin{figure}
\centering
\includegraphics{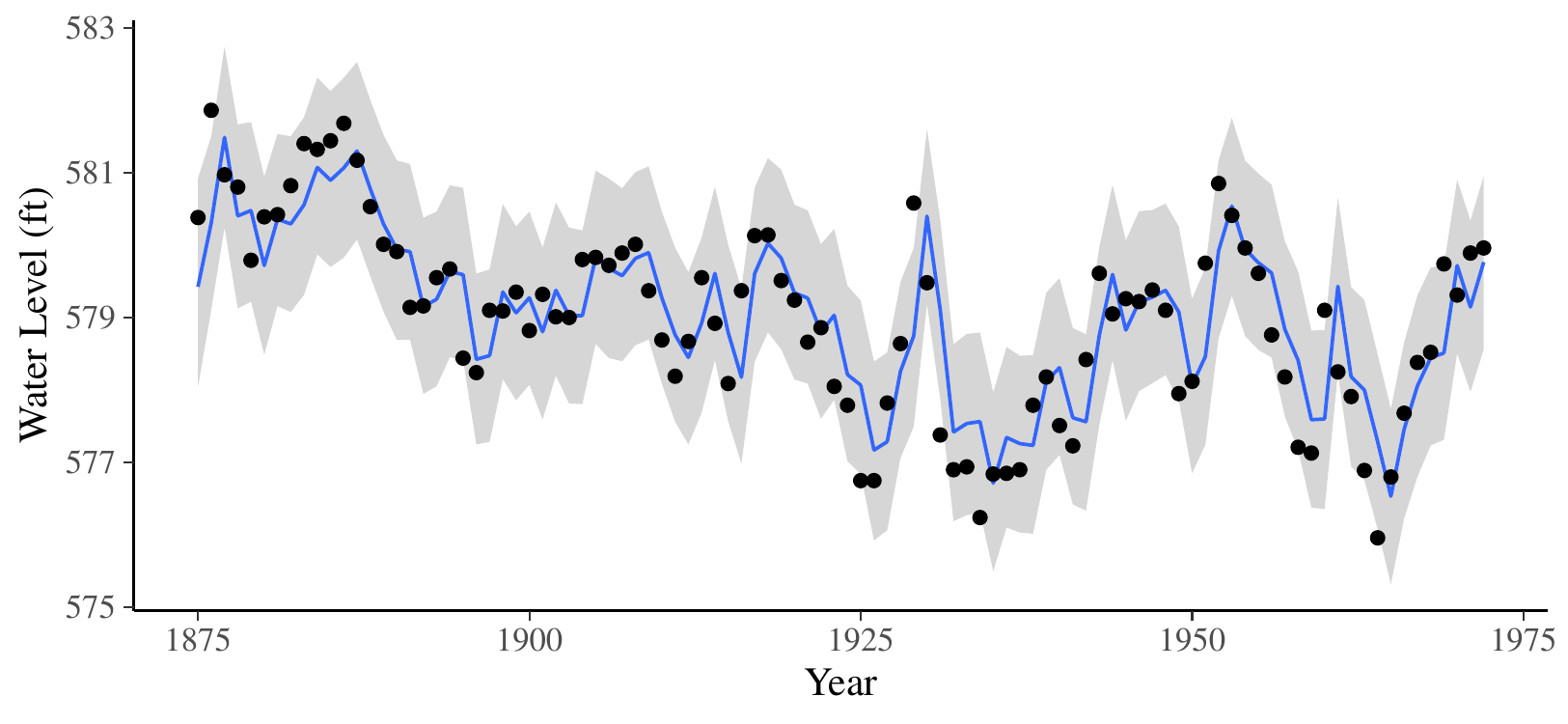}
\caption{\label{fig:lake-huron}Water Level in Lake Huron (1875-1972). Black points are observed data. The blue line represents mean predictions of an AR(4) model with 90\% prediction intervals shown in gray.}
\end{figure}

To illustrate the application of PSIS-LFO-CV for estimating expected \(M\)-SAP
performance, we will fit a model for 98 annual measurements of the water level
(in feet) of \href{https://en.wikipedia.org/wiki/Lake_Huron}{Lake Huron} from the
years 1875--1972. This data set is found in the \emph{datasets} R package, which is
installed automatically with R \citep{R2018}. The time series shows rather strong
autocorrelation and some downward trend towards lower water levels for later points
in time. Figure \ref{fig:lake-huron} shows the observed time series of water
levels as well as predictions from a fitted AR(4) model.

Based on this data and model, we will illustrate the use of PSIS-LFO-CV to
provide estimates of \(1\)-SAP and \(4\)-SAP when leaving out all future values. To
allow for reasonable predictions, we will require at least \(L = 20\) historical
observations (20 years) to make predictions. Further, we set a threshold of
\(\tau =\) 0.7 for the Pareto \(k\) estimates that indicate when refitting
becomes necessary. Our fully reproducible analysis of this case study can be
found on GitHub (\url{https://github.com/paul-buerkner/LFO-CV-paper}).

\begin{figure}
\centering
\includegraphics{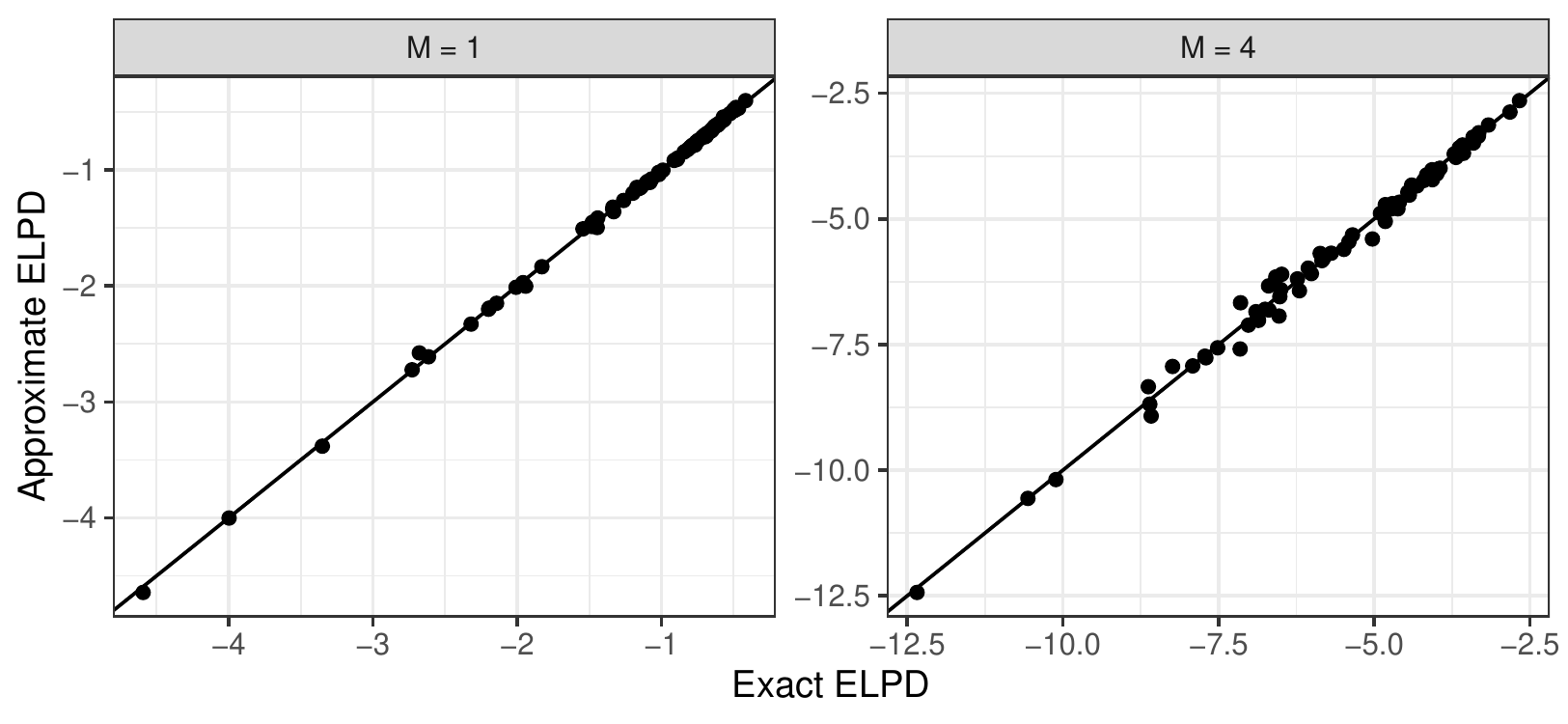}
\caption{\label{fig:lh-pw-elpd}Pointwise exact vs.~PSIS-approximated ELPD contributions for 1-SAP (left) and 4-SAP (right) for the Lake Huron model. A threshold of \(\tau = 0.7\) was used for the Pareto \(k\) estimates. \(M\) is the number of predicted future observations.}
\end{figure}

\begin{figure}
\centering
\includegraphics{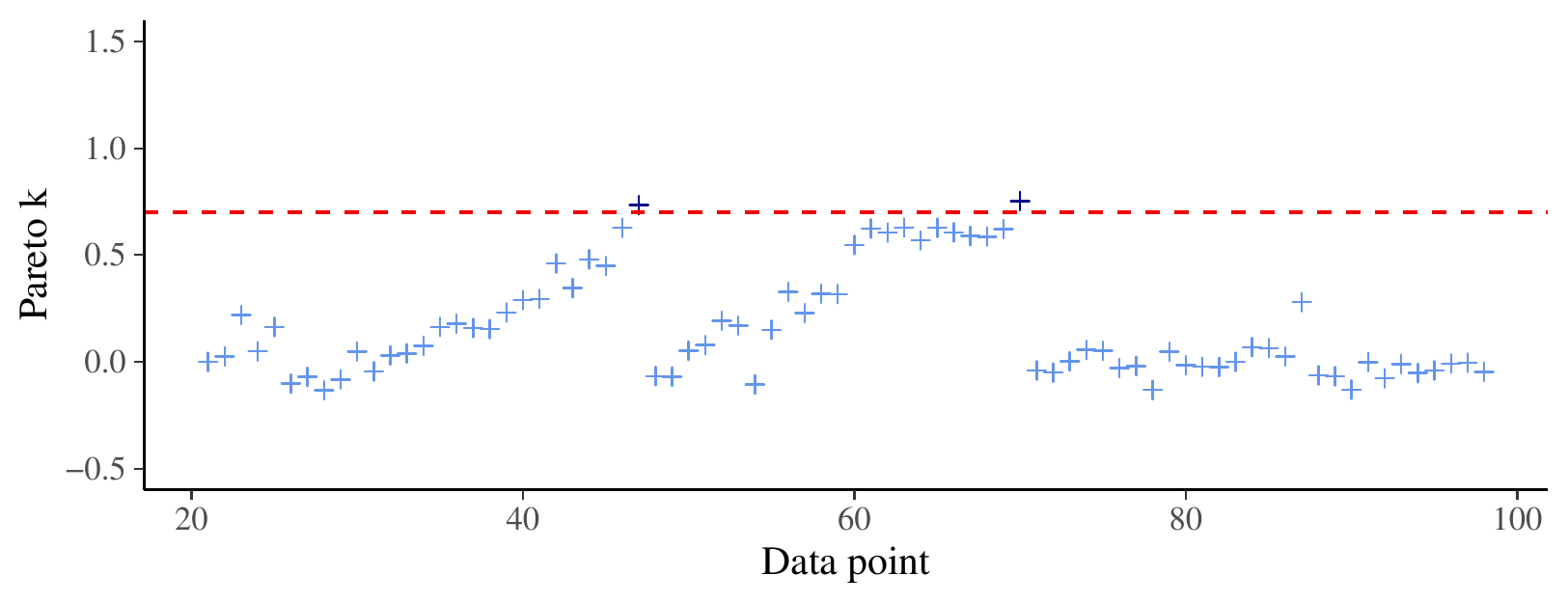}
\caption{\label{fig:lh-pareto-k}Pareto \(k\) estimates for PSIS-LFO-CV of the Lake Huron model. The dotted red line indicates the threshold at which the refitting was necessary.}
\end{figure}

We start by computing exact and PSIS-approximated LFO-CV of 1-SAP. The computed
ELPD values are
\({\rm ELPD}_{\rm exact} =\) -93.48 and
\({\rm ELPD}_{\rm approx} =\) -93.62, which are
almost identical. Not only is the overall ELPD estimated accurately but so are
all of the pointwise ELPD contributions (see the left panel of Figure
\ref{fig:lh-pw-elpd}). In comparison, PSIS-LOO-CV returns
\({\rm ELPD}_{\rm loo} =\) -88.9,
overestimating the predictive performance and as suggested by our simulation
results for stationary autoregressive models (see fourth row of Figure
\ref{fig:1sap}). Plotting the Pareto \(k\) estimates reveals that the model had
to be refit 3 times, out of a total of \(N - L =\) 78 predicted
observations (see Figure \ref{fig:lh-pareto-k}). On average, this means one
refit every 26.0 observations, which implies a drastic
speed increase compared to exact LFO-CV.

Performing LFO-CV for 4-SAP, we obtained \({\rm ELPD}_{\rm exact} =\)
-411.41 and \({\rm ELPD}_{\rm approx} =\)
-412.78, which are again very similar.
In general, as \(M\) increases, the approximation will tend to become more variable
around the true value in absolute ELPD units because the ELPD increment of each
observation will be based on more and more observations (see also Section
\ref{simulations}). For this example, we see some considerable differences in
the pointwise ELPD contributions of specific observations which were hard to
predict accurately by the model (see the right panel of Figure
\ref{fig:lh-pw-elpd}). This is to be expected because predicting \(M\) steps ahead
using an AR model will yield highly uncertain predictions if most of the
autocorrelation happens at lags smaller than \(M\) (see also the bottom rows in
Figure \ref{fig:4sap}). For such a model, it may be ill-advised to evaluate
predictions too far into the future, at least when using the approximate methods
presented in this paper. Since, for a constant threshold \(\tau\), the importance
weights are the same independent of \(M\), the Pareto \(k\) estimates are the same
for \(4\)-SAP and \(1\)-SAP.

\hypertarget{case-CB}{%
\subsection{Annual date of the cherry blossoms in Japan}\label{case-CB}}

The cherry blossom in Japan is a famous natural phenomenon occurring once every
year during spring. As the climate changes so does the annual date of the cherry
blossom \citep{aono2008, aono2010}. The most complete reconstruction available to
date contains data between 801 AD and 2015 AD
\citep{aono2008, aono2010} and is available online
(\url{http://atmenv.envi.osakafu-u.ac.jp/aono/kyophenotemp4/}).

\begin{figure}
\centering
\includegraphics{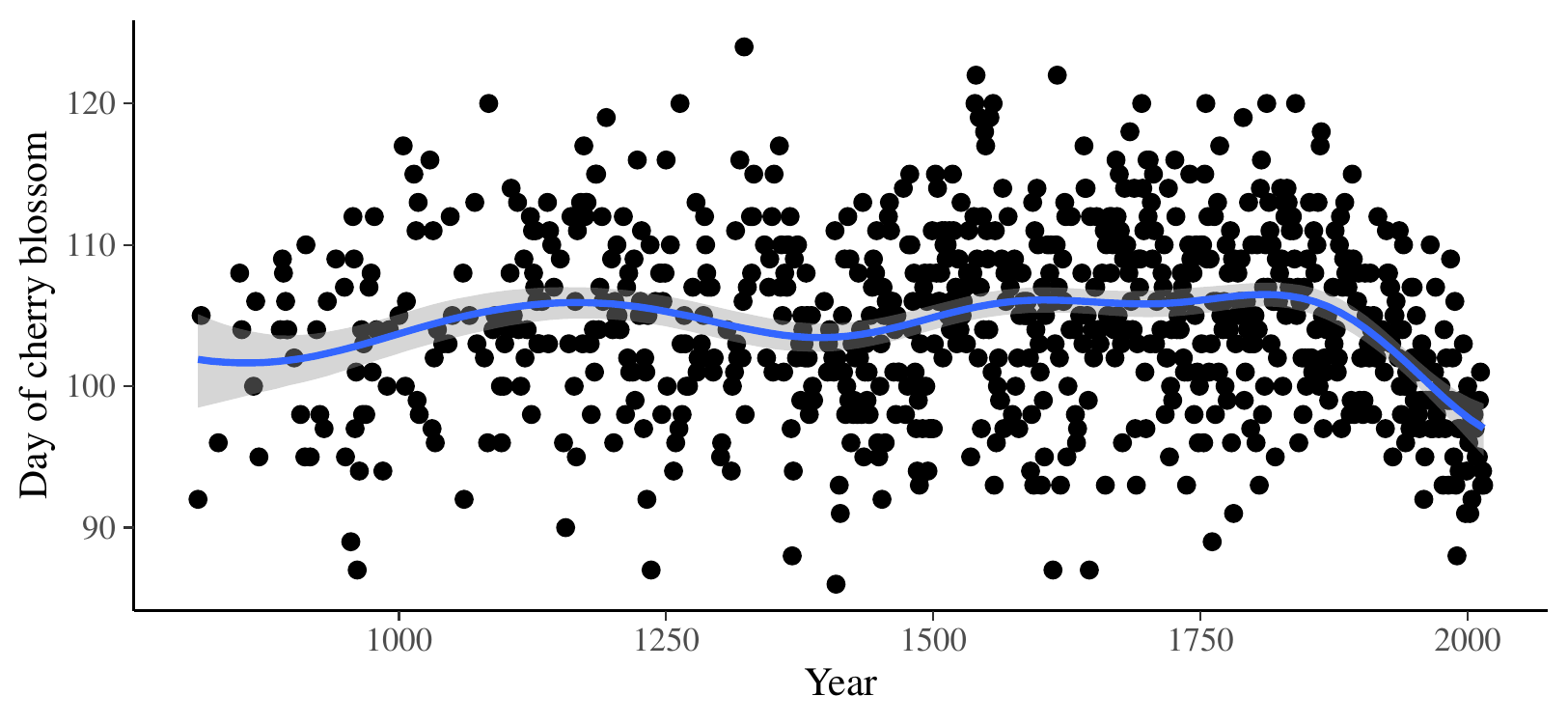}
\caption{\label{fig:cherry-blossom}Day of the cherry blossom in Japan (812-2015). Black points are observed data. The blue line represents mean predictions of a thin-plate spline model with 90\% regression intervals shown in gray.}
\end{figure}

In this case study, we will predict the annual date of the cherry
blossom using an approximate Gaussian process model \citep{solin2014, RiutortMayol2019} to
provide flexible non-linear smoothing of the time series. A visualisation of
both the data and the fitted model is provided in Figure
\ref{fig:cherry-blossom}. While the time series appears rather stable across
earlier centuries, with substantial variation across consecutive years, there
are some clearly visible trends in the data. Particularly in more recent years,
the cherry blossom has tended to happen much earlier than before, which may
be a consequence of changes in the climate \citep{aono2008, aono2010}.

Based on this data and model, we will illustrate the use of PSIS-LFO-CV to
provide estimates of \(1\)-SAP and \(4\)-SAP leaving out all future values. To allow
for reasonable predictions of future values, we will require at least \(L = 100\)
historical observations (100 years) to make predictions. Further, we set a
threshold of \(\tau =\) 0.7 for the Pareto \(k\) estimates to determine when
refitting becomes necessary. Our fully reproducible analysis of this case study
can be found on GitHub (\url{https://github.com/paul-buerkner/LFO-CV-paper}).

\begin{figure}
\centering
\includegraphics{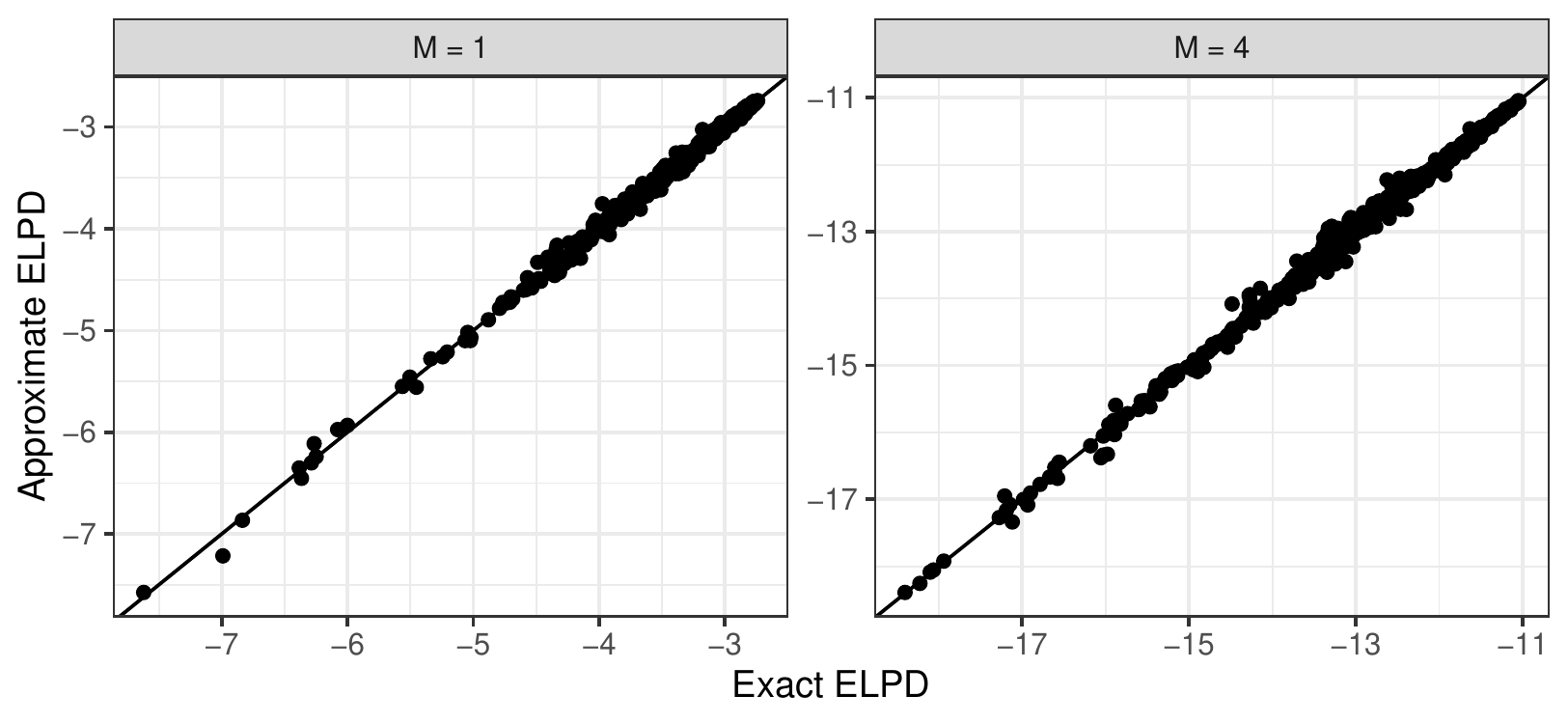}
\caption{\label{fig:cb-pw-elpd}Pointwise exact vs.~PSIS-approximated ELPD contributions of 1-SAP (left) and 4-SAP (right) for the cherry blossom model. A threshold of \(\tau = 0.7\) was used for the Pareto \(k\) estimates. \(M\) is the number of predicted future observations.}
\end{figure}

\begin{figure}
\centering
\includegraphics{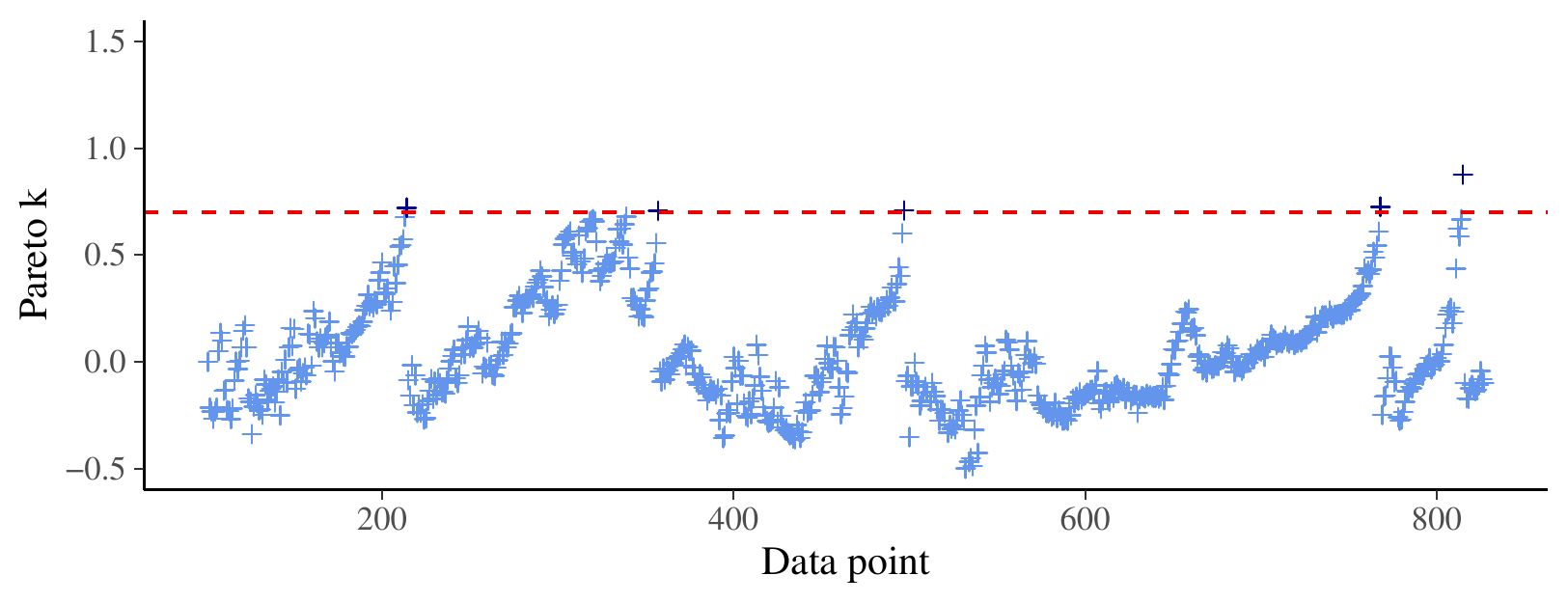}
\caption{\label{fig:cb-pareto-k}Pareto \(k\) estimates for PSIS-LFO-CV of the cherry blossom model. The dotted red line indicates the threshold at which the refitting was necessary.}
\end{figure}

We start by computing exact and PSIS-approximated LFO-CV of 1-SAP. We compute
\({\rm ELPD}_{\rm exact} =\) -2345.7 and
\({\rm ELPD}_{\rm approx} =\) -2344.9,
which are highly similar. As shown in the left panel of Figure
\ref{fig:cb-pw-elpd}, the pointwise ELPD contributions are highly accurate,
with no outliers. The approximation has worked well for all observations.
PSIS-LFO-CV performs much better than PSIS-LOO-CV
(\({\rm ELPD}_{\rm approx} =\) -2340.3), which
overestimates the predictive performance. Plotting the Pareto \(k\) estimates
reveals that the model had to be refit 6 times, out of a total of \(N - L =\) 727 predicted observations (see Figure \ref{fig:cb-pareto-k}). On
average, this means one refit every 121.2 observations,
which implies a drastic speed increase as compared to exact LFO-CV.

Performing LFO-CV of 4-SAP, we compute \({\rm ELPD}_{\rm exact} =\)
-9348.3 and \({\rm ELPD}_{\rm approx} =\)
-9345.5, which are again similar but not as close
as the corresponding 1-SAP results. This is to be expected as the uncertainty of
PSIS-LFO-CV increases for increasing \(M\) (see Section \ref{simulations}). As
displayed in the right panel of Figure \ref{fig:cb-pw-elpd}, the pointwise ELPD
contributions are highly accurate in most cases, with a few small outliers in
both directions. For constant threshold \(\tau\), the importance weights are the
same independent of \(M\), so the Pareto \(k\) estimates are the same for \(4\)-SAP
and \(1\)-SAP.

\hypertarget{discussion}{%
\section{Conclusion}\label{discussion}}

We proposed, evaluated, and demonstrated PSIS-LFO-CV, a method for approximating
cross-validation of Bayesian time series models. PSIS-LFO-CV is intended to be
used when the prediction task is predicting future values based solely on past
values, in which case leave-one-out cross-validation is inappropriate.
Within the set of such prediction tasks, we can choose the number \(M\) of future
observations to be predicted. For a set of common time series models, we
established via simulations that PSIS-LFO-CV is an unbiased approximation
of exact LFO-CV if we choose the threshold \(\tau\) of the Pareto \(k\) estimates to
not be larger than \(\tau = 0.7\). That is, PSIS-LFO-CV does not require a smaller
(stricter) threshold than PSIS-LOO-CV to achieve satistfactory accuracy.

By nature of the approximated \(M\)-step-ahead predictions, the computation time
of PSIS-LFO-CV still increases linearily with the number of observations \(N\).
However, in our numerical experiments, we were able to reduce to computation
time by a factor of roughly 25 to 100 compared to exact LFO-CV, which is
enough to make LFO-CV realistic for many applications.

A limitation of our current approach is that the uncertainty in the approximate
LFO-CV estimates is hard to quantify. There are at least three types of
uncertainty which could be considered here. First, there is uncertainty induced
by approximating exact LFO-CV using (Pareto smoothed) importance sampling. Based
on theoretical considerations of the approximation and numerical experiments
both presented in \citet{vehtari2019psis}, any PSIS approximation will be very close to
the exact value as long as the Pareto \(k\) diagnostic does not exceed the
threshold of \(0.7\), which we used as the refit criterion in our approximate
LFO-CV approach. Second, there is uncertainty caused by finite amounts of data.
For 1-step-ahead predictions, we can use an analogous approach to what is done in
approximate LOO-CV by computing the standard error across the pointwise
estimates for each observation \citep{vehtari2017loo}. More generally, for
\(M\)-step-ahead predictions, we can compute the standard error by using every
\(M\)th value which are then independent. Third, there is uncertainty induced by
the finite number of posterior draws but this uncertainty tends to be negligable
with just a few thousand draws compared to the second source of uncertainty
\citep{vehtari2017loo}. Investigating uncertainty measures for (approximate) LFO-CV
in more detail is left for future research.

Lastly, we want to briefly note that LFO-CV can also be used to compute marginal
likelihoods. Using basic rules of conditional probability, we can factor the
log marginal likelihood as

\begin{equation}
\log p(y) = \sum_{i=1}^N \log p(y_i \,|\, y_{1:(i-1)}).
\end{equation}

This is exactly the ELPD of 1-SAP if we set \(L = 0\), that is if we
choose to predict \emph{all} observations using their respective past (the very
first observation is only predicted from the prior). As such, marginal
likelihoods may be approximated using PSIS-LFO-CV. Although this approach is
unlikely to be more efficient than methods specialized for computing marginal
likelihoods \citep[e.g., bridge sampling;][]{meng1996, meng2002, gronau2017}, it may
be a noteworthy option if for some reason other methods fail.

\hypertarget{acknowledgments}{%
\section{Acknowledgments}\label{acknowledgments}}

We thank Daniel Simpson, Shira Mitchell, M\r{a}ns Magnusson, and anonymous
reviewers for helpful comments and discussions on earlier versions of this
paper. We acknowledge the Academy of Finland (grants 298742, 313122) as well as
the Technology Industries of Finland Centennial Foundation (grant 70007503;
Artificial Intelligence forResearch and Development) for partial support of this
work. We also acknowledge the computational resources provided by the Aalto
Science-IT project.

\newpage

\hypertarget{appendix}{%
\section*{Appendix}\label{appendix}}
\addcontentsline{toc}{section}{Appendix}

\hypertarget{appendix-a-pseudo-code-for-psis-lfo-cv}{%
\subsection*{Appendix A: Pseudo code for PSIS LFO-CV}\label{appendix-a-pseudo-code-for-psis-lfo-cv}}
\addcontentsline{toc}{subsection}{Appendix A: Pseudo code for PSIS LFO-CV}

The R flavored pseudo code below provides a description of the proposed
PSIS-LFO-CV algorithm when leaving out all future values. See \url{https://github.com/paul-buerkner/LFO-CV-paper} for the actual R code.

\begin{Shaded}
\begin{Highlighting}[]
\CommentTok{# Approximate Leave-Future-Out Cross-Validation (LFO-CV)}
\CommentTok{# Arguments:}
\CommentTok{#   model: the fitted time series model based on the complete data}
\CommentTok{#   data: the complete data set}
\CommentTok{#   M: number of steps to be predicted into the future}
\CommentTok{#   L: minimal number of observations necessary to make predictions}
\CommentTok{#   tau: threshold of the Pareto-k-values}
\CommentTok{# Returns:}
\CommentTok{#   PSIS approximated ELPD value of LFO-CV}
\NormalTok{PSIS_LFO_CV =}\StringTok{ }\ControlFlowTok{function}\NormalTok{(model, data, M, L, tau) \{}
\NormalTok{  N =}\StringTok{ }\KeywordTok{number_of_rows}\NormalTok{(data)}
\NormalTok{  S =}\StringTok{ }\KeywordTok{number_of_draws}\NormalTok{(model)}
\NormalTok{  out =}\StringTok{ }\KeywordTok{vector}\NormalTok{(}\DataTypeTok{length =}\NormalTok{ N)}
  \CommentTok{# refit the model using the first L observations}
\NormalTok{  i_star =}\StringTok{ }\NormalTok{L}
\NormalTok{  model_star =}\StringTok{ }\KeywordTok{update}\NormalTok{(model, }\DataTypeTok{data =}\NormalTok{ data[}\DecValTok{1}\OperatorTok{:}\NormalTok{L, ])}
\NormalTok{  out[L] =}\StringTok{ }\KeywordTok{exact_ELPD}\NormalTok{(model_star, }\DataTypeTok{data =}\NormalTok{ data[(L }\OperatorTok{+}\StringTok{ }\DecValTok{1}\NormalTok{)}\OperatorTok{:}\NormalTok{(L }\OperatorTok{+}\StringTok{ }\NormalTok{M), ])}
  \CommentTok{# loop over all observations at which to perform predictions}
  \ControlFlowTok{for}\NormalTok{ (i }\ControlFlowTok{in}\NormalTok{ (L }\OperatorTok{+}\StringTok{ }\DecValTok{1}\NormalTok{)}\OperatorTok{:}\NormalTok{(N }\OperatorTok{-}\StringTok{ }\NormalTok{M)) \{}
\NormalTok{    PSIS_object =}\StringTok{ }\KeywordTok{PSIS}\NormalTok{(model_star, }\DataTypeTok{data =}\NormalTok{ data[(i_star }\OperatorTok{+}\StringTok{ }\DecValTok{1}\NormalTok{)}\OperatorTok{:}\NormalTok{i , ])}
\NormalTok{    k =}\StringTok{ }\KeywordTok{pareto_k_values}\NormalTok{(PSIS_object)}
    \ControlFlowTok{if}\NormalTok{ (k }\OperatorTok{>}\StringTok{ }\NormalTok{tau) \{}
      \CommentTok{# refitting the model is necessary}
\NormalTok{      i_star =}\StringTok{ }\NormalTok{i}
\NormalTok{      model_star =}\StringTok{ }\KeywordTok{update}\NormalTok{(model_star, }\DataTypeTok{data =}\NormalTok{ data[}\DecValTok{1}\OperatorTok{:}\NormalTok{i, ])}
\NormalTok{      out[i] =}\StringTok{ }\KeywordTok{exact_ELPD}\NormalTok{(model_star, }\DataTypeTok{data =}\NormalTok{ data[(i }\OperatorTok{+}\StringTok{ }\DecValTok{1}\NormalTok{)}\OperatorTok{:}\NormalTok{(i }\OperatorTok{+}\StringTok{ }\NormalTok{M), ])}
\NormalTok{    \} }\ControlFlowTok{else}\NormalTok{ \{}
      \CommentTok{# PSIS approximation is possible}
\NormalTok{      log_PSIS_weights =}\StringTok{ }\KeywordTok{log_weights}\NormalTok{(PSIS_object)}
\NormalTok{      out[i] =}\StringTok{ }\KeywordTok{approx_ELPD}\NormalTok{(model_star, }\DataTypeTok{data =}\NormalTok{ data[(i }\OperatorTok{+}\StringTok{ }\DecValTok{1}\NormalTok{)}\OperatorTok{:}\NormalTok{(i }\OperatorTok{+}\StringTok{ }\NormalTok{M), ],}
                           \DataTypeTok{log_weights =}\NormalTok{ log_PSIS_weights)}
\NormalTok{    \}}
\NormalTok{  \}}
  \KeywordTok{return}\NormalTok{(}\KeywordTok{sum}\NormalTok{(out))}
\NormalTok{\}}
\end{Highlighting}
\end{Shaded}

\hypertarget{appendix-b-backward-psis-lfo-cv}{%
\subsection*{Appendix B: Backward PSIS-LFO-CV}\label{appendix-b-backward-psis-lfo-cv}}
\addcontentsline{toc}{subsection}{Appendix B: Backward PSIS-LFO-CV}

Instead of moving forward in time, that is, starting our predictions from the
\(L\)th observation, we may also move backwards, a procedure to which we will
refer to as backward PSIS-LFO-CV. Starting with \(i = N - M\), we approximate each
\(p(y_{i+1:M} \,|\, y_{1:i})\) via

\begin{equation}
 p(y_{i+1:M} \,|\, y_{1:i}) \approx
   \frac{ \sum_{s=1}^S w_i^{(s)}\, p(y_{i+1:M} \,|\, y_{1:i}, \theta^{(s)})}
        { \sum_{s=1}^S w_i^{(s)}},
\end{equation}

where \(w_i^{(s)}\) are the PSIS weights and \(\theta^{(s)} = \theta^{(s)}_{1:i^\star}\) are draws from the posterior distribution based on the
first \(1:i^\star\) observations. In backward LFO-CV, we start using the model
based on \emph{all} observations, that is, set \(i^\star = N\). To obtain \(w_i^{(s)}\),
we first compute the raw importance ratios

\begin{equation}
r_i^{(s)} = r_i(\theta^{(s)}) = 
\frac{f_{1:i}(\theta^{(s)})}{f_{1:i^\star}(\theta^{(s)})} 
\propto \frac{
p(\theta^{(s)}) \prod_{j \in 1:i} p(y_j \,|\, y_{1:(j-1)}, \theta^{(s)})
}{
p(\theta^{(s)}) \prod_{j \in 1:i^\star} p(y_j \,|\, y_{1:(j-1)}, \theta^{(s)})
}
= \frac{1}{\prod_{j \in (i+1):i^\star} p(y_j \,|\, y_{1:(j-1)}, \theta^{(s)})},
\end{equation}

and then stabilize them using PSIS as described in Section \ref{psis}. The
function \(f_{1:i}\) denotes the posterior distribution based on the first \(i\)
observations, that is, \(f_{1:i} = p(\theta \,|\, y_{1:i})\), with \(f_{1:i^\star}\)
defined analogously. The index set \((i + 1):i^\star\) indicates all observations
which are part of the data for the actually fitted model \(f_{1:i^\star}\) but not
for the model \(f_{1:i}\) whose predictive performance we are trying to
approximate. The proportional statement arises from the fact that we ignore the
normalizing constants \(p(y_{1:i})\) and \(p(y_{1:i^\star})\) of the compared
posteriors, which leads to a self-normalized variant of PSIS \citep[c.f.][]{vehtari2017loo}. This approach to computing importance ratios is a
generalization of the approach used in PSIS-LOO-CV, where only a single
observation is left out at a time.

\begin{figure}
\centering
\includegraphics{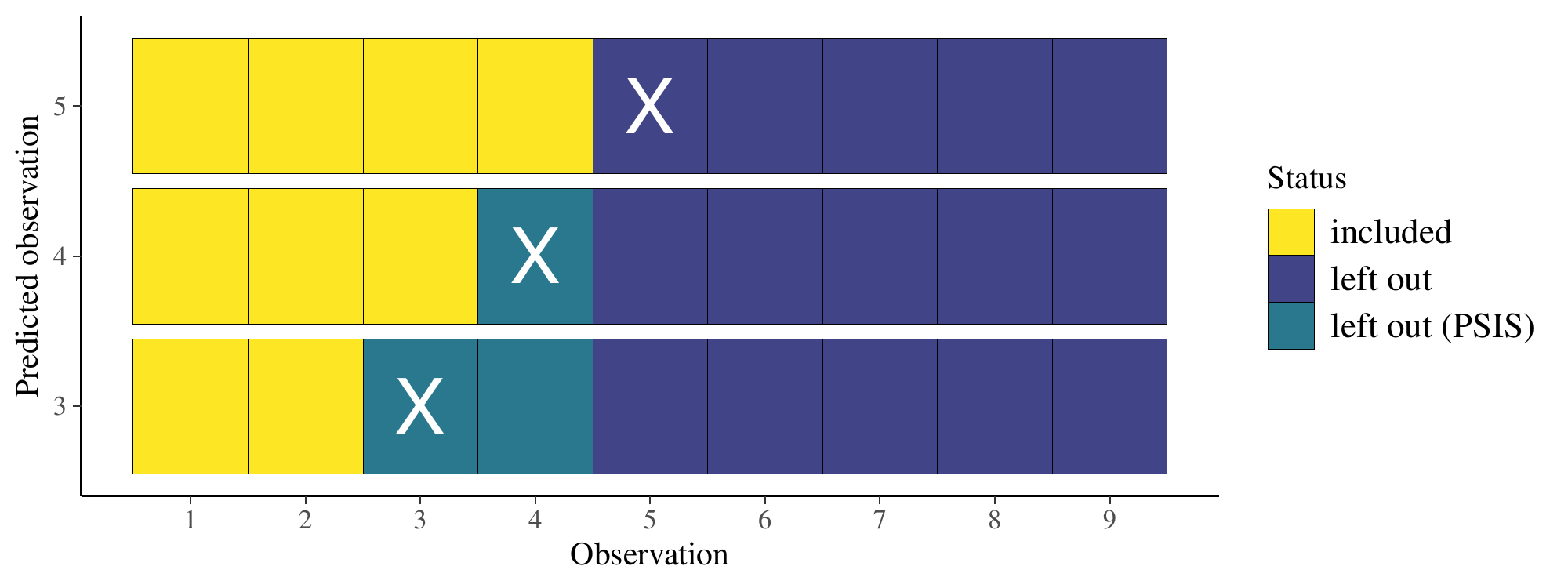}
\caption{\label{fig:vis-msap-bw}Visualisation of approximate one-step-ahead predictions using backward PSIS-LFO-CV. Predicted observations are indicated by \textbf{X}. In the shown example, the model was last refit at the \(i^\star = 4\)th observation.}
\end{figure}

Starting from \(i = N - M\), we gradually \emph{decrease} \(i\) by \(1\) (i.e., we move
backwards in time) and repeat the process. At some observation \(i\), the
variability of the importance ratios \(r_i^{(s)}\) will become too large and
importance sampling fails. We will refer to this particular value of \(i\) as
\(i^\star_1\). To identify the value of \(i^\star_1\), we check for which value of
\(i\) does the estimated shape parameter \(k\) of the generalized Pareto
distribution first cross a certain threshold \(\tau\) \citep{vehtari2019psis}. Only
then do we refit the model using only observations up to \(i^\star_1\) by setting
\(\theta^{(s)} = \theta^{(s)}_{1:i^\star_1}\) as well as \(i^\star = i^\star_1\)
and restarting the process. An illustration of this procedure is shown
in Figure \ref{fig:vis-msap-bw}. In some cases we may only need to refit once
and in other cases we will find a value \(i^\star_2\) that requires a second
refitting, maybe an \(i^\star_3\) that requires a third refitting, and so on. We
repeat the refitting as many times as is required (only if \(k > \tau\)) until we
arrive at \(i = L\). Recall that \(L\) is the minimum number of observations we have
deemed acceptable for making predictions.

In forward PSIS-LFO-CV, we have seen a threshold of \(\tau = 0.7\) to be
sufficient for achieving satisfactory accuracy. For backward PSIS-LFO-CV, \(\tau\)
likely has to be smaller. More precisely, we can expect an appropriate threshold
for the backward mode to be somewhere between \(0.5 \leq \tau \leq 0.7\). It is
unlikely to be as high as the \(\tau = 0.7\) default used for PSIS-LOO-CV because
there will be more dependence in the errors in the case of backward PSIS-LFO-CV.
If there is a large error when leaving out the \(i\)th observation, then there is
likely to also be a large error when leaving out observations \(i, i-1, i-2, \ldots\) until a refit is performed. This means that highly influential
observations (ones with a large \(k\) estimate) are likely to have stronger
effects on the total estimate for backward LFO-CV than for LOO-CV.

\begin{figure}
\centering
\includegraphics{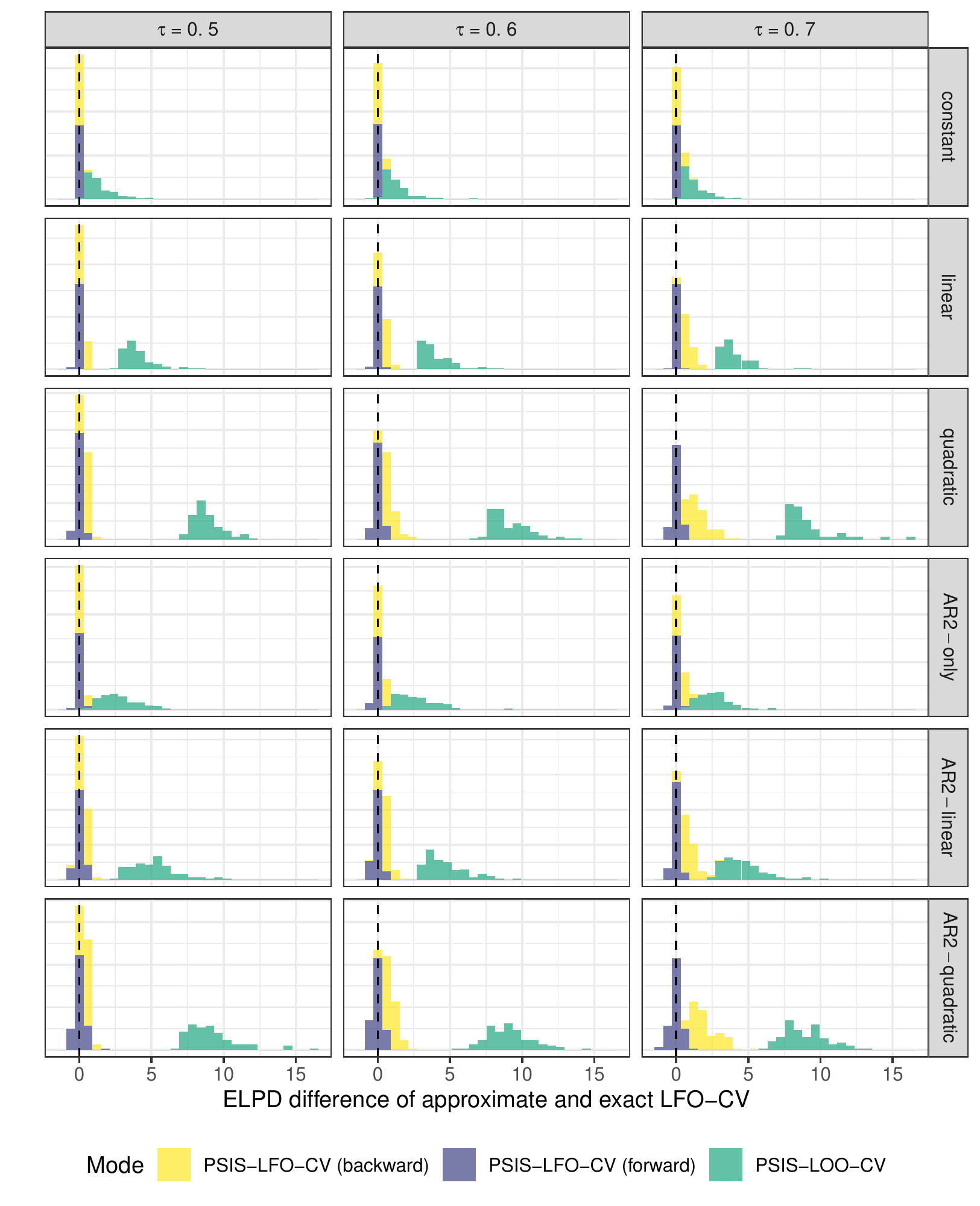}
\caption{\label{fig:1sap-both}Simulation results of 1-step-ahead predictions for both forward and backward PSIS-LFO-CV. Histograms are based on 100 simulation trials of time series with \(N = 200\) observations requiring at least \(L = 25\) observations to make predictions. The black dashed lines indicates the exact LFO-CV result.}
\end{figure}

\begin{figure}
\centering
\includegraphics{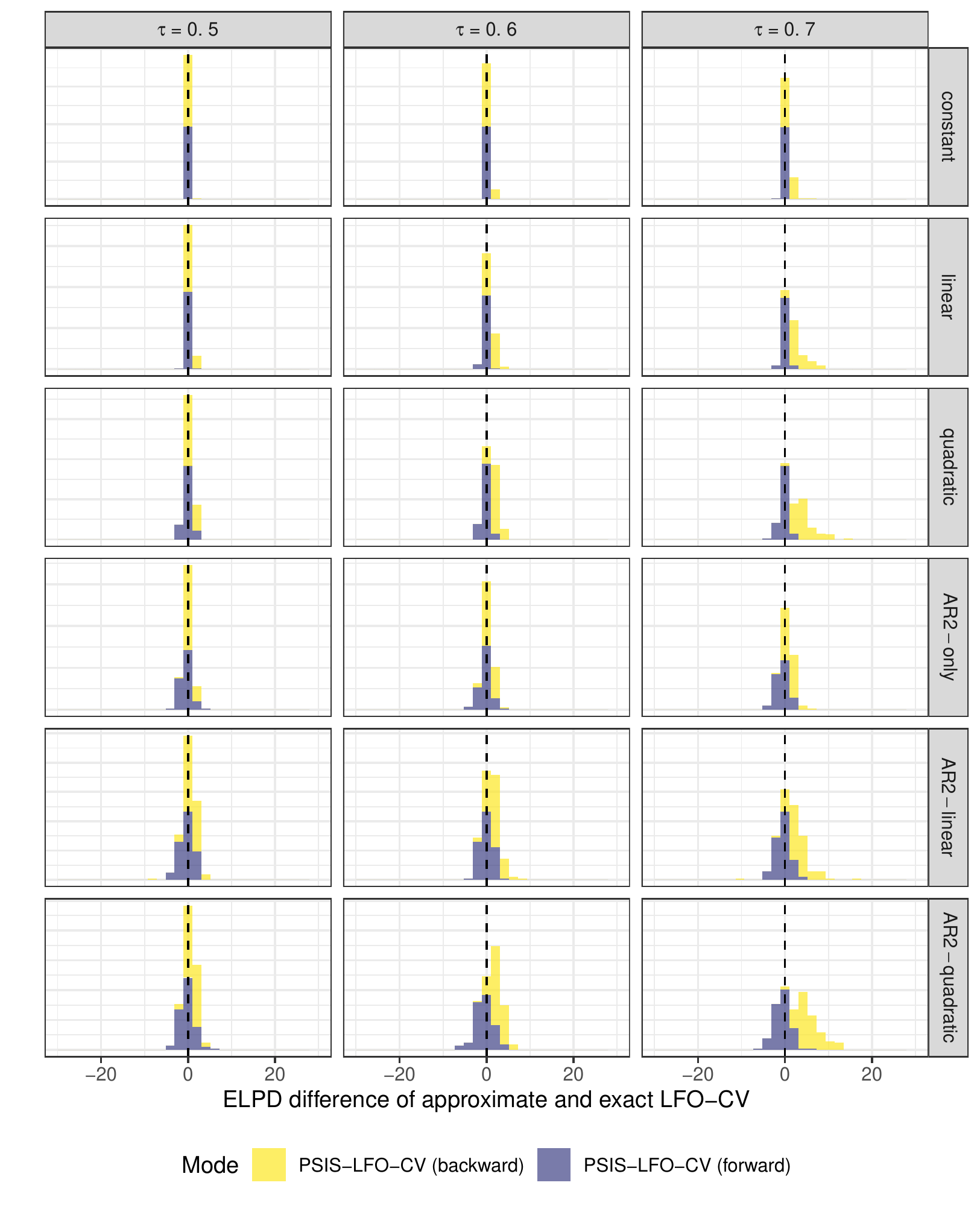}
\caption{\label{fig:4sap-both}Simulation results of 4-step-ahead predictions for both forward and backward PSIS-LFO-CV. Histograms are based on 100 simulation trials of time series with \(N = 200\) observations requiring at least \(L = 25\) observations to make predictions. The black dashed lines indicates the exact LFO-CV result.}
\end{figure}

The simulation results comparing backward to forward PSIS-LFO-CV can be found in
Figure \ref{fig:1sap-both} for \(1\)-SAP and in Figure \ref{fig:4sap-both} for
\(4\)-SAP. As visible in both figures, backward PSIS-LFO-CV requires a lower
\(\tau\) threshold than forward PSIS-LFO-CV in order to be accurate (\(\tau = 0.6\)
vs.~\(\tau = 0.7\)). Otherwise, it may have a small positive bias. Further, as can
be seen in Table \ref{tab:refits-both}, backward PSIS-LFO-CV requires
considerably more refits then forward PSIS-LFO-CV. Together, this indicates
that, in expectation, backward PSIS-LFO-CV is inferior to forward PSIS-LFO-CV.

\begin{table}

\caption{\label{tab:refits-both}Mean proportions of required refits for both forward and backward PSIS-LFO-CV.}
\centering
\begin{threeparttable}
\begin{tabular}[t]{llrrrrrrr}
\toprule
Mode & M & $\tau$ & constant & linear & quadratic & AR2-only & AR2-linear & AR2-quadratic\\
\midrule
backward & 1 & 0.5 & 0.03 & 0.08 & 0.17 & 0.04 & 0.09 & 0.18\\
 &  & 0.6 & 0.02 & 0.06 & 0.12 & 0.03 & 0.06 & 0.12\\
 &  & 0.7 & 0.01 & 0.04 & 0.09 & 0.02 & 0.04 & 0.08\\
 & 4 & 0.5 & 0.03 & 0.08 & 0.17 & 0.04 & 0.09 & 0.18\\
 &  & 0.6 & 0.02 & 0.06 & 0.12 & 0.03 & 0.06 & 0.12\\
\addlinespace
 &  & 0.7 & 0.01 & 0.04 & 0.09 & 0.02 & 0.04 & 0.09\\
forward & 1 & 0.5 & 0.01 & 0.01 & 0.02 & 0.01 & 0.02 & 0.03\\
 &  & 0.6 & 0.01 & 0.01 & 0.02 & 0.01 & 0.02 & 0.02\\
 &  & 0.7 & 0.01 & 0.01 & 0.02 & 0.01 & 0.01 & 0.02\\
 & 4 & 0.5 & 0.01 & 0.01 & 0.02 & 0.01 & 0.02 & 0.03\\
\addlinespace
 &  & 0.6 & 0.01 & 0.01 & 0.02 & 0.01 & 0.02 & 0.02\\
 &  & 0.7 & 0.01 & 0.01 & 0.02 & 0.01 & 0.01 & 0.02\\
\bottomrule
\end{tabular}
\begin{tablenotes}
\item Note: Results are based on 100 simulation trials of time series with $N = 200$ observations requiring at least $L = 25$ observations to make predictions. Abbreviations: $\tau$ = threshold of the Pareto $k$ estimates; $M$ = number of predicted future observations.
\end{tablenotes}
\end{threeparttable}
\end{table}

We may even combine forward and backward mode PSIS-LFO-CV in the following way.
First, we start with forward mode until a refit becomes necessary, say at
observation \(i^\star\). Then, we apply backward mode on the basis of the refitted
model and perform multiple proposal importance sampling \citep{veach1995, he2014} to
obtain the ELPD values of the observations \(i^\star - 1, i^\star - 2, \ldots\)
from the mixture of the forward and backward distributions. We do this until the
backward mode requires a refit at which point we stop the process and continue
with forward mode at observation \(i^\star\). This algorithm requires exactly as
many refits as the forward mode while potentially increasing accuracy for those
observations for which the pointwise ELPD contribution was computed via both
forward and backward mode PSIS-LFO-CV. In the present paper, we did not
investigate the possibility of multiple importance sampling in more detail, but
it could be a promising extention to be studied in the future.

\hypertarget{appendix-c-psis-lfo-cv-for-the-rmse}{%
\subsection*{Appendix C: PSIS-LFO-CV for the RMSE}\label{appendix-c-psis-lfo-cv-for-the-rmse}}
\addcontentsline{toc}{subsection}{Appendix C: PSIS-LFO-CV for the RMSE}

We may also use other measures of predictive performance than the ELPD,
for instance the RMSE. For a scalar response \(y\) and corresponding vector
\(\hat{y}\) of a total of \(S\) posterior predictions \(\hat{y}^{(s)}\), the RSME
is defined as

\begin{equation}
\text{RMSE}(y, \hat{y}) = \frac{1}{S} \sum_{s=1}^S (\hat{y}^{(s)} - y)^2.
\end{equation}

If we predict multiple responses in the future (i.e., perform \(M\)-SAP with
\(M > 1\)), we simply sum the RMSE over all those responses.
When approximating the RMSE via PSIS, we use the (Pareto smoothed)
importance weights \(w^{(s)}\) (see Section \ref{psis-MSAP}) to estimate

\begin{equation}
\text{RMSE}(y, \hat{y}) \approx
  \frac{\sum_{s=1}^S w^{(s)} (\hat{y}^{(s)} - y)^2}{\sum_{s=1}^S w^{(s)}}.
\end{equation}

The remaining computations are analogous to using the ELPD as a measure of
predictive performance in LFO-CV and so we do not spell out the details here.
The code we provide on GitHub (\url{https://github.com/paul-buerkner/LFO-CV-paper}) is
modularized and also has an implementation of the (approximate) RMSE for LFO-CV.

\begin{figure}
\centering
\includegraphics{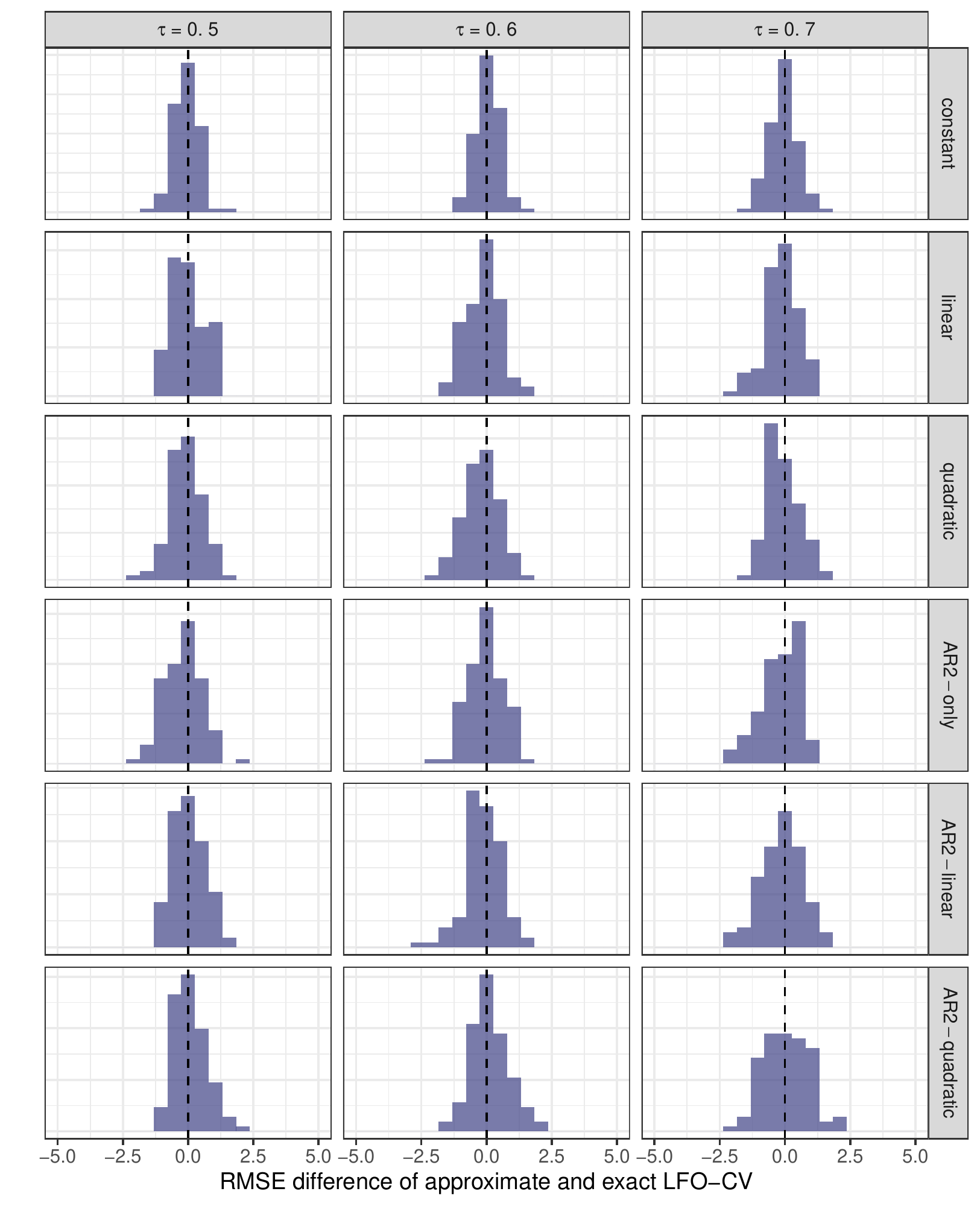}
\caption{\label{fig:1sap-rmse}Simulation results of 1-step-ahead predictions using the RMSE as measure of predictive accuracy. Histograms are based on 100 simulation trials of time series with \(N = 200\) observations requiring at least \(L = 25\) observations to make predictions. The black dashed lines indicates the exact LFO-CV result.}
\end{figure}

\begin{figure}
\centering
\includegraphics{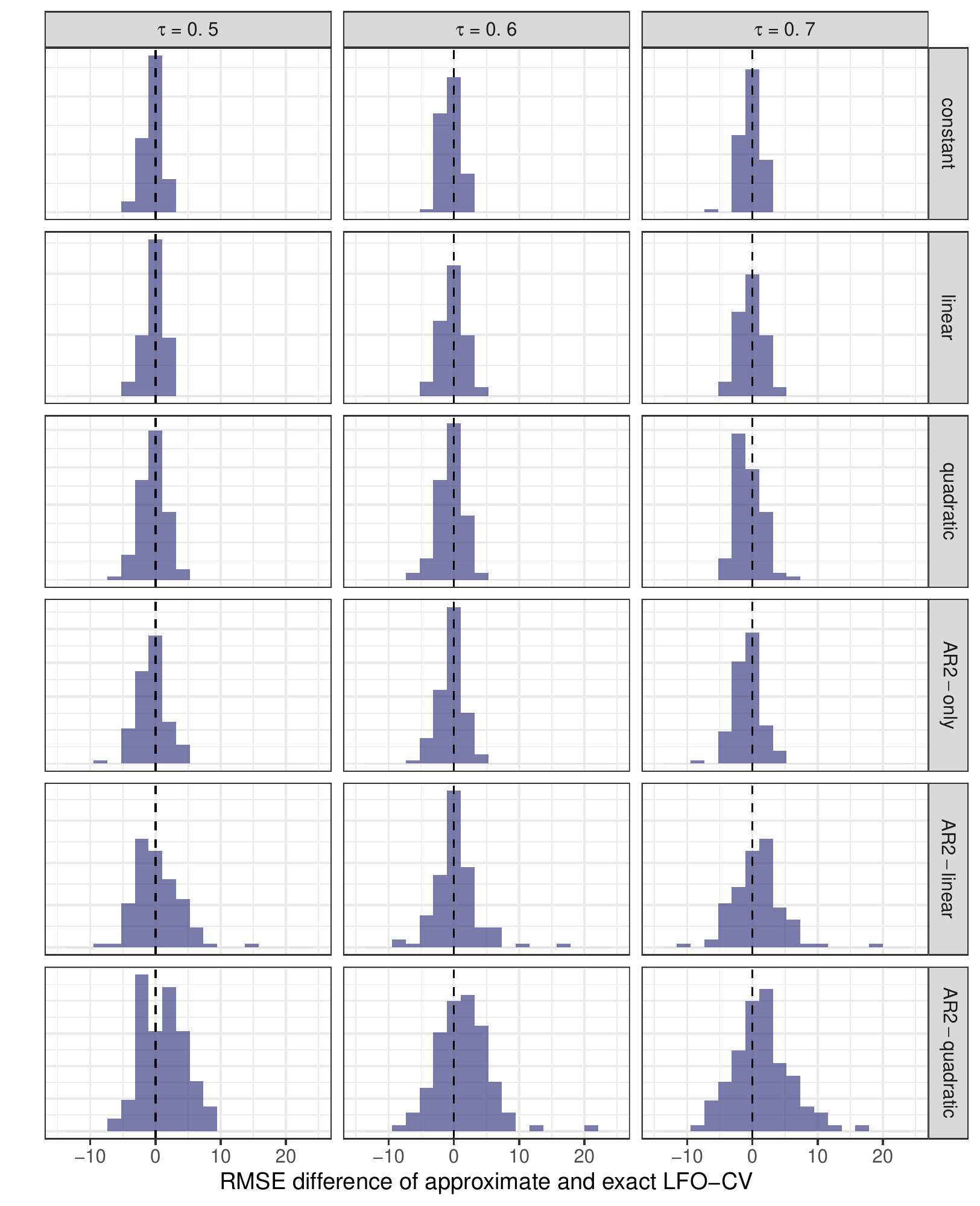}
\caption{\label{fig:4sap-rmse}Simulation results of 4-step-ahead predictions using the RMSE as measure of predictive accuracy. Histograms are based on 100 simulation trials of time series with \(N = 200\) observations requiring at least \(L = 25\) observations to make predictions. The black dashed lines indicates the exact LFO-CV result.}
\end{figure}

Results of the 1-SAP and 4-SAP RMSE simulations are visualized in Figure
\ref{fig:1sap-rmse} and \ref{fig:4sap-rmse}, respectively. It is clearly
visible that the the accuracy of the PSIS RMSE approximation is nearly
independent of the threshold \(\tau\) when \(\tau\) is within the interval
\([0.5,0.7]\) motivated in \ref{psis} \citep[this would not be the case if \(\tau\) was
allowed to be larger;][]{vehtari2019psis}. For all conditions, the PSIS-LFO-CV
approximation is highly accurate, that is, both approximately unbiased and low
in variance around the corresponding exact LFO-CV RMSE value (represented by the
dashed line in Figure \ref{fig:1sap}). Taken together, these simulations
indicate that PSIS-LFO-CV not only works well with the ELPD but also with the
RMSE.

\end{document}